\documentclass[review, 11pt]{elsarticle}

\usepackage{amsmath}
\usepackage{amsfonts}
\usepackage[noend]{algpseudocode}
\usepackage{amssymb}
\usepackage[thmmarks,amsmath]{ntheorem}
\usepackage{mathrsfs}
\usepackage{fancyhdr}
\usepackage{algorithm}
\usepackage[english]{babel}
\usepackage{t1enc}
\usepackage[utf8]{inputenc} 
\usepackage{ccaption}
\usepackage{psfrag}
\usepackage{color}
\usepackage{epic}
\usepackage{eepic}
\usepackage{enumerate}
\usepackage{array}
\usepackage[titletoc,title]{appendix}
\usepackage{picins}
\usepackage{pslatex}
\usepackage{ulem}
\usepackage{comment}
\usepackage{multirow}

\usepackage{graphicx}
\usepackage{rotating}
\usepackage{chronology}
\usepackage{scrextend}
\usepackage[caption=false,font=footnotesize]{subfig}
\usepackage[margin=3cm]{geometry}

 \algnewcommand{\And}{\textbf{and}}
  \algnewcommand{\Or}{\textbf{or}}

\usepackage{lineno,hyperref}
\modulolinenumbers[5]

\journal{Applied Soft Computing}

\usepackage{numcompress}

\usepackage{pdfpages}
\usepackage{array}
\begin{document}
\newpage
\begin{frontmatter}
\setcounter{page}{1}
\title{
The Value of Big Data for Credit Scoring: \\Enhancing Financial Inclusion using Mobile Phone Data and Social Network Analytics}
\cortext[cor1]{Corresponding author}
\author[ru]{Mar\'{i}a \'{O}skarsd\'{o}ttir}\corref{cor1}
\ead{mariaoskars@ru.is}
\author[wu]{Cristi\'{a}n Bravo}
\ead{cBravoRo@uwo.ca}
\author[arg]{Carlos Sarraute}
\ead{charles@grandata.com}
\author[feb]{Jan Vanthienen}
\ead{jan.vanthienen@kuleuven.be}
\author[feb,sth]{Bart Baesens}
\ead{bart.baesens@kuleuven.be}
\address[ru]{Department of Computer Science, Reykjavik University, Iceland}
\address[wu]{Department of Statistical and Actuarial Sciences, University of Western Ontario}
\address[feb]{Dept. of Decision Sciences and Information Management, KU Leuven, Belgium.}
\address[sth]{Dept. of Decision Analytics and Risk, University of Southampton, United Kingdom.}
\address[arg]{Grandata Labs, San Francisco, California, USA.}

\begin{abstract}
Credit scoring is without a doubt one of the oldest applications of analytics.
In recent years, a multitude of sophisticated classification techniques have been developed to improve the statistical performance of credit scoring models.
Instead of focusing on the techniques themselves, this paper leverages alternative data sources to enhance both statistical and economic model performance. 
The study demonstrates how including call networks, in the context of positive credit information, as a new Big Data source has added value in terms of profit by applying a profit measure and profit-based feature selection.
A unique combination of datasets, including call-detail records, credit and debit account information of customers is used to create scorecards for credit card applicants.
Call-detail records are used to build call networks and advanced social network analytics techniques are applied to propagate influence from prior defaulters throughout the network to produce influence scores.
The results show that combining call-detail records with traditional data in credit scoring models significantly increases their performance when measured in AUC.
In terms of profit, the best model is the one built with only calling behavior features. 
In addition, the calling behavior features are the most predictive in other models, both in terms of statistical and economic performance.
The results have an impact in terms of ethical use of call-detail records, regulatory implications, financial inclusion, as well as data sharing and privacy.

 \end{abstract}
\begin{keyword}
Credit Scoring\sep Social Network Analysis\sep Profit Measure \sep Mobile Phone Data 
\end{keyword}
\date{}
\end{frontmatter}

\section{Introduction}\label{secIntroduction}
Credit scoring is undoubtedly one of the oldest applications of analytics where lenders and financial institutions perform statistical analysis to assess the creditworthiness of potential borrowers to help them decide whether or not to grant credit \citep{thomas2000survey}.
Fair Isaac was founded in 1956 as one of the first analytical companies offering retail credit scoring services in the US.  
Its well-known FICO score (ranging between 300 and 850) has been used as a key decision instrument by financial institutions, insurers, utilities companies and even employers \citep{scheule2016credit}.  
The first corporate credit scoring models date back to the late sixties with Edward Altman developing his well-known z-score model for bankruptcy prediction, which is still used to this day in Bloomberg reports as a default risk benchmark \citep{altman1968financial}.  
Originally, these models were built using limited data--consisting of only a few hundred observations--and were based on simple classification techniques such as linear programming, discriminant analysis and logistic regression, which is the current industry standard given its high interpretability \citep{scheule2016credit}.  
The importance of these retail and corporate credit scoring models further increased due to various regulatory compliance guidelines such as the Basel Accords and IFRS 9 which clearly stipulate the inputs and outputs of a credit scoring model together with how these models can be used to calculate provisions and capital buffers.  

The emergence of more sophisticated classification techniques such as neural networks, support vector machines and random forests led to various extensive benchmarking studies aimed at improving credit scoring models in terms of their statistical performance (e.g., in terms of area under the ROC curve or classification accuracy) \citep{baesens2003benchmarking, lessmann2015benchmarking}.  
Many of these studies concluded that traditional credit scoring models based on, e.g., simple logistic regression models, performed very well and newer classification techniques could only offer marginal performance gains.  
In other words, research on developing high-performing credit scoring models has more or less stalled.  
We believe the best investment in better credit scoring models is not to turn the attention to newer classification techniques but to leverage innovative Big Data sources instead.  

While these new sources of data present the opportunity to profile potential borrowers using a wider representation of behavior, they also present an ethical challenge. Mobile phone data, e.g., in the form of call-detail records (CDR), allows constructing a very detailed social network, and using this information to profile repayment behavior can be seen as unfair to borrowers that could be punished for their mobile cell phone behavior. Recently, the use of \emph{positive information} has been put forward as a necessary source of data that should be included in scoring models \citep{worldbank2011}. Positive information is defined as all information that represents the good financial behavior, providing a clearer definition of the factors that make a good borrower. \citet{barron2003value} show, for example, that not using positive information leads to a decrease of up to 47.5\% in credit availability. 

This paper introduces mobile phone data as a new Big Data source for credit scoring and shows that while it is a powerful source of information, it should be used strictly in a positive framework to increase the access to financing to borrowers who would otherwise be out of options until a much later stage. To motivate the use of this information in financial institutions, its potential is studied in both statistical and profit terms.

Big Data is typically defined in terms of its 5 Vs: Volume, Variety, Velocity, Veracity and Value.  
Recent special issues of Information Systems Research \citep{agarwal2014editorial} and MIS Quarterly \citep{baesens2014transformational} indicate the explosion of interest in Big Data within the IS community.  
The use of mobile phone data for credit scoring is a fitting example of this since it comes in huge volumes (Volume), has not been explored before (Variety), is generated on a continuous daily basis (Velocity) and is usually stored using a well-defined call-detail record log format (Veracity).  
In this paper, its Value is also quantified by focusing both on its statistical performance (e.g., in terms of area under the ROC curve) and on its bottom line impact in terms of profit.
Additionally, an evaluation of the qualitative performance of the data in terms of positive information for enhanced financial inclusion is provided.
This study is based on a unique data set combining banking, sociodemographic and CDR data.
CDR are logs of all phone calls between the customers of a telecommunications provider (telco), see Table \ref{T:CDR}. 
More specifically, the data set includes all CDR of the bank's customers, the CDR of the people they are in contact with and the banking history of these customers.  
Overall, it adds up to a year and a half of banking history of over two million bank customers and the calling activity of almost 90 million unique phone numbers spanning five months. 
This unique combination of data gives the opportunity to explore the potential of enriching traditional credit scoring models with social network effects reflecting calling behavior.  
The three key research questions are:
\begin{itemize}
\item[Q1]
What is the added value (in terms of both AUC and profit) of including call data for credit scoring?
\item[Q2]
Can call data replace traditional data used for credit scoring?
\item[Q3]
How does default behavior propagate in the call network?
\end{itemize}
To the best of our knowledge, these questions have not been researched before.  
Each of the questions will be answered from both a statistical as well as a profit perspective, which is another key contribution of this paper. 
Furthermore, the implications for financial inclusion are evaluated.

The impact of this research is manifold.  
A successful application of boosting the performance of credit scoring models using call data would improve credit decision-making and pricing.  
The insights could also facilitate access to credit for borrowers with little or no credit history. 
This is the case for young borrowers, lenders exploring new markets or in developing countries with young credit markets.
In all these cases, the borrowers are not expected to have a credit history, but they do have mobile phone records.  
Knowing how default behavior propagates in a call network also has regulatory implications.  
For example, the Basel Accords try to capture default correlation in order to better protect a financial institution against unexpected losses \citep{scheule2016credit}.  
The research can shed new light on how default behavior is correlated.  
This could lead to better provisioning and capital buffering strategies, thereby improving the resilience of the financial system against shocks and macroeconomic downturns.
Knowing how default behavior propagates in a call network also has other regulatory implications. 
If CDR data is indeed useful for credit prediction, then banks and credit bureaus have a strong economic incentive to collaborate with telecommunications companies to share data in order to perform this type of analyses. 

In the next section, a review on the literature on Big Data in credit scoring as well as previous research on call networks is provided.
In section \ref{sec:methodology}, the theoretical background and methodology applied in the case study is described, with the experimental setup detailed in section \ref{sec:experimentaldesign}.
The results are presented in section \ref{sec:results}, followed by a discussion on their various implications in sections \ref{sec:discuss} and \ref{sec:impact}.
The paper concludes with a summary of the contributions and discussion on possibilities for future work.

\section{Related Work}\label{sec:relatedwork}

Many analytical modeling exercises start from a flat data set, build a predictive model for a target measure of interest (e.g., churn, fraud, default) and evaluate it on an independent out-of-sample data set.  
An assumption which is (oftentimes) tacitly made is that the data is independent and identically distributed. 
Recent research questioned this assumption and analyzed how customers can influence each other through the different social networks that connect them \citep{sundararajan2013research}.  
Various types of social behavior can be observed. 
One is homophily, which states that people have a strong tendency to associate with others whom they perceive as being similar to themselves in some way. 
Social influence occurs when people's behavior is affected by others with whom they interact \citep{newman2010networks, lee2016friend}.  
Some of the social behavior can also be attributed to other (e.g., external) confounding factors \citep{aral2014tie}. 
The idea of network learning is to embed social behavior patterns in the predictive models to successfully leverage the impact of joint customer actions \citep{macskassy2007classification}.
A key input to any social network learning exercise is the network itself, which consist of nodes and edges.  
In certain settings, the definition of these networks is relatively straightforward.  
As an example, consider churn prediction in telco where the network can obviously be constructed based upon data stored in the CDR.  
Earlier research found significant social network effects for predicting churn in telco \citep{verbeke2014social}.  
Another example is credit card fraud detection where a network can be defined by connecting credit cards to merchants.  
Also in this setting, strong social network effects have been found \citep{van2015apate}.  

In credit scoring, there is a firm belief amongst both researchers and practitioners that default behavior of borrowers is correlated.  
To illustrate this, the Basel Accord models default correlation by means of an asset correlation term, which is set to $15\%$ for residential mortgages and $4\%$ for qualifying revolving exposures.  
However, both these numbers have been set in a rather arbitrary way, or based upon some empirical but not published procedure \citep{gordy2003risk}. 
This interdependency has been proven to be a significant factor amongst small and medium-sized enterprises \citep{calabrese2017birds}.
One of the key challenges in understanding network effects or default propagation in credit scoring concerns the definition of the network itself.  
Preliminary attempts have been made to build networks between customers in online peer-to-peer lending.  
For example, \citet{lin2013judging} illustrated that online friendships with non-defaulters increases the credit score. 
These findings were also confirmed by \citet{freedman2014information}, with an additional caution that online ties on their own may not reveal true information about creditworthiness and may also be manipulated \citep{wei2015credit}.  
\citet{de2018does} developed credit scoring models for microfinance using social media network information extracted from Facebook accounts.  
Their results suggest that explicit networks of friends who interact are more predictive than of friends who do not, but implicit networks of people with similar behavior are better than both explicit friendship networks.  
In industry, social networks are already being exploited to assess creditworthiness, by technology companies such as Lenddo, that make use of social media connections to analyse people's default risk \citep{kharif2016no}.

More recently, the interest in using call networks as a new Big Data source for credit scoring has gained traction, e.g., with \citet{wei2015credit} formulating the potential value of credit scores obtained with networks--for example, based on social media or calls--and how strategic tie-formation might affect these scores.
Although especially interesting in relation to the Chinese government's plan for a social credit system 
\citep{chin2016china}, the study is only theoretical and is missing an important empirical evaluation of the proposed models \citep{wei2015credit}.
Moreover, recent press coverage on specialized smartphone applications that evaluate people's creditworthiness using the huge amount of data generated by their handsets indicates the potential of call networks as an alternative data source for credit scoring \citep{dwoskin2015lending, kharif2016no}.  
Most of these studies have focused on the use of social networks in the context of social media, or have discussed the potential of CDR-induced social networks in credit scoring.

The literature on the analysis of CDR is rich \citep{naboulsi2016large}. 
The idea of using CDR data for credit scoring stems from the fact that the way people use their phone is assumed to be a good proxy for their lifestyle and economic activity.
Previous research confirmed that using CDR data to build call networks by linking together individuals who are in contact with each other, results in social networks that can be used in both descriptive and predictive studies on age, gender, ethnicity, language, economic factors, geography, urbanization, and epidemics \citep{blondel2008fast,
onnela2011geographic,wesolowski2012quantifying, sarraute2014study,leo2016socioeconomic}.  
For example, \citet{leo2016socioeconomic} confirm the presence of homophily in terms of economic behavior using call networks.  
More specifically, they show that wealth and debt are unevenly distributed and that people are better connected with those that share their socioeconomic class. 
Furthermore, \citet{haenlein2011social} investigated the distribution of customer revenue within a call network and demonstrated that high revenue customers are primarily related to other high revenue customers and the same for low revenue customers.  

\section{Methodology}\label{sec:methodology}
This paper contributes to the literature by investigating the use of CDR data for credit scoring in terms of value.
Here, the proposed methodology for extracting appropriate information from the CDR data by means of social networks and influence propagation is detailed.
Furthermore, techniques for evaluating model and feature performance in terms of profit are presented. 

\subsection{Call Networks: Featurization and Propagation}\label{subsec:callNetworks}
\begin{table}
\centering
\caption{An example of a CDR log. In the actual dataset the phone numbers are encrypted. \label{T:CDR}}
\scriptsize{
{\begin{tabular}{llrrr}
\hline
Call Start Date&Call Start Time&Call Duration (sec)& From Number& To Number\\ \hline
01MAY2017& 14:51:14& 715 &(202) 555-0116& (701) 555-0191 \\
02MAY2017& 14:34:37& 29& (803) 555-0129& (202) 555-0116 \\
01MAY2017&20:34:14& 9 &(803) 555-0117& (406) 555-0137\\
02MAY2017& 20:03:38& 89& (701) 555-0148& (803) 555-0129 \\ \hline
\end{tabular}}}
\end{table}
A call network is a network where the nodes $\mathcal{V}=\{v_1,\dots v_n\}$ are people present in a CDR log. 
These logs are kept for billing purposes and include information about every phone call made by the customers of a telecommunications operator, including the encrypted phone numbers of the customers that made and received the phone call as well as timing and length.
An example of such a log can be seen in Table \ref{T:CDR}.  
Information from CDR about time and duration of phone calls (or text messages) can be used to connect the people in the network to create the edges, $e_{i,j}\in\mathcal{E}$.
The edges are either undirected, such as when two customers share a phone call but it is irrelevant which customer made the call; or directed, in which case we distinguish between outgoing and incoming edges (i.e., all phone calls made by and received by a person, respectively). 
The edges are represented by an $n$ by $n$ binary matrix, called adjacency matrix $A$, where a non-zero entry denotes an existing edge between node $v_i$ and $v_j$ in an undirected network and from/to $v_i$ to/from $v_j$ in a directed network with outgoing/incoming edges.
The edges can also carry weights to indicate the intensity of the relationship between two people, for example the number or duration of phone calls they share in a given time period.
The weights are denoted by the weight matrix $W=(w_{i,j})$, where $w_{i,j}\in\mathbb{R}^{+}\cup\{0\}$.
The first order neighborhood of a node $v_i$ is the collection of nodes $v_j$ that share an edge with $v_i$, that is 
\[
N^1_i=\{v_j|e_{i,j}\in\mathcal{E}, j=1,\dots,n\}.
\]
In some networks, the nodes can be labelled, or assigned to a class that is later used in a predictive analytics framework. 
In this application, there are two types of labels.
The first type of label regards default, in which case the customers in the call network belong to one of two classes:
they are either defaulters, who have been in arrears for more than 90 days within a twelve month period (bad customers); or they are non-defaulters (good customers). \footnote{We use the Basel definition of default. \citep{scheule2016credit}}
When building a credit scoring model, the goal is to assign one of these two classes to each customer of interest and it is the target variable of the classification problem in this study.
In the call network there are also customers who, during the timespan of the network, have run into payment arrears for one or two months in addition to defaulters with three months of payment arrears.
For clarity, all these customers are referred to as delinquent customers and it is the second type of label.
The delinquent customers have the possibility to influence others in the network to also run into payment arrears--also referred to as default influence--and they are used when generating features as explained below.

\begin{figure}
\centering
{\includegraphics[scale=0.35]{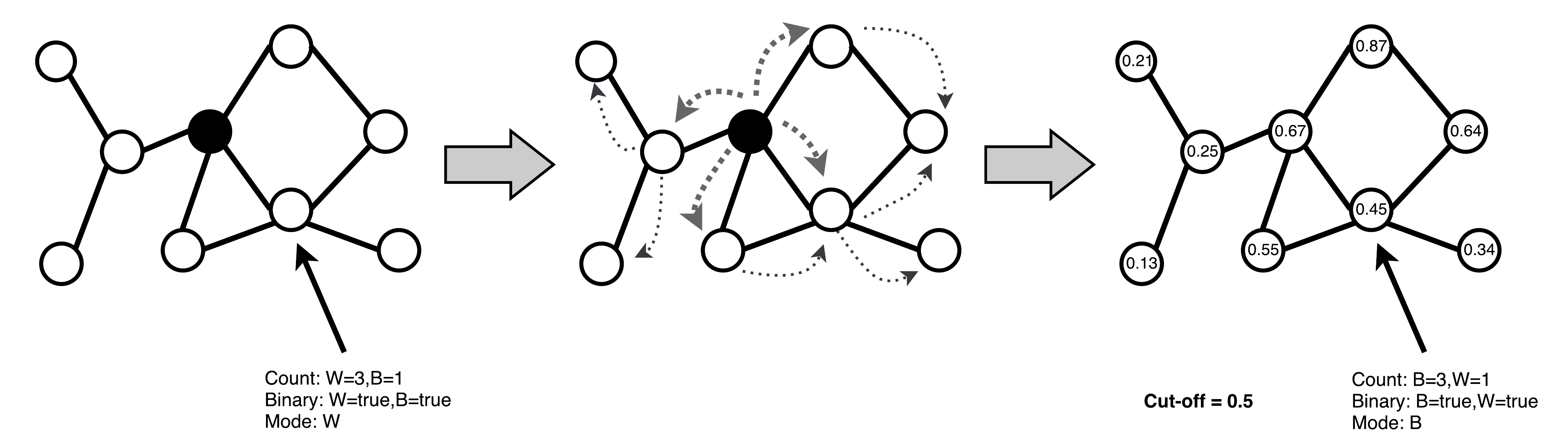}}
\caption{The figure demonstrates the computation of link-based measures, before and after a propagation method is applied to the network.
The figure on the left shows a network with one black node and eight white nodes. 
The link-based features of the node to which the arrow points are summarized, where B means black and W means white.  
The figure in the middle demonstrates the application of the propagation method with the resulting exposure scores shown for each node in the figure on the right.
After the exposure scores have been computed, a cut-off point is set at 0.5 and nodes with a score that is higher than the threshold are labelled black (B), and white (W) otherwise.
Subsequently, link-based exposure features are extracted for the node to which the arrow points.\label{fig:propagation}}
\end{figure}

In order to use the information that is contained in the call networks for building credit scoring models, network features are extracted for each node in the network by aggregating information about its position within the network and connectivity to other nodes.
As in similar studies, a distinction is made between direct network features, which are derived from the node's first order neighborhood, and indirect network features that take into account the whole network structure \citep{van2016gotcha}. 
As stated earlier, the aim is to study how delinquent customers may influence others with whom they are connected.
Therefore, by assuming there is prior knowledge about some delinquent customers in the network (i.e., having a subset of nodes with known labels) that knowledge can be incorporated in the network features by exploiting social ties.
To this end, both direct and indirect network features are extracted as illustrated in Figure \ref{fig:propagation}. 
The direct network features represent the presence and number of delinquent customers in a node's first order neighborhood.
They are easy to extract and provide a representative overview of people's social connections \citep{lu2003link}.
However, the influence of payment arrears is likely to reach further than just the first order neighborhood.
This effect is modeled using two distinct propagation methods that have been effective in previous research and are designed to simulate real-life behavior: Personalized PageRank (PR) and Spreading Activation (SPA).
The results of both methods are exposure scores which are categorized as indirect network features.
Although other propagation methods exist, such as Gibbs sampling and relaxation labelling, these were not applied here because they have been shown to be less effective for prediction in call networks \citep{oskarsdottir2017social}, are less scalable and as such did not fulfil the requirements of this study. 
The features resulting from these three approaches will be used as input features when building credit scoring models, but first a more detailed explanation is provided.
\subsubsection{Link-Based Features}
\citet{lu2003link} presented a framework for inferring labels for nodes in a network based on labels of neighboring nodes.
They defined three features that can be extracted from the neighborhood of a node: count-link, mode-link and binary-link. 
These represent, respectively, the frequency of classes in the neighborhood, their mode, and a binary indicator for each class.
Futhermore, using a logistic regression model, \citet{lu2003link} showed that these features are very predictive for the class of the node itself.

Extraction of link-based features is based on the presence of delinquent customers with varying number of payment arrears.
\subsubsection{Personalized PageRank}
The propagation method Personalized PageRank (PR) was developed for search engines (e.g., Google) to rank webpages while also taking into account an initial source of information, such as frequently visited web pages \citep{page1999pagerank} but can also be used for different kinds of linked data \citep{van2015apate,van2016gotcha}.
For the nodes in a network with weight matrix $W$, the method iteratively computes exposure scores $\xi_{k+1}$ based on the exposure scores in the node's neighborhood $\xi_{k}$ and a random jump to other nodes in the network, determined by the information source $z$ -- also called restart vector --  using the equation
\[
\xi_{k+1}=\alpha W \xi_k +(1-\alpha)z,
\]
where $1-\alpha$, the damping factor, denotes the probability of a random jump and $k$ is the iteration step.
As a result of the initial information source, exposure scores of nodes closer to the source nodes are higher. 
Here, the delinquent customers are the information source.
\subsubsection{Spreading Activation}
The propagation method Spreading Activation (SPA) originates from cognitive psychology and simulates how information, or energy, spreads through the network from a set of source nodes.
It is used to model a `word-of-mouth' scenario, where influence--in this case from delinquent customers--spreads through the network.
`Word-of-mouth' has been shown to be effective in social networks \citep{dasgupta2008social,backiel2015combining}.
Before the method begins, a set of active nodes  $V^A\subset \mathcal{V}$ possesses the energy $E^0(V_A)$.
In each step $k$ of this iterative method, a part $d$ of an active node's energy $E^k(V_A)$ is spread to the nodes in its neighborhood while the rest of the energy remains.
The part that is transferred, is distributed according to the relative weights of the links to neighboring nodes, expressed by the transfer function
\[
E_{transfer}=\frac{d\cdot w_{i,j}}{\sum_{w_{i,s}\in N_i^1} w_{i,s}} E^k(V^A_i).
\]
The method stops when no more nodes are being affected and the changes in energy of the already affected nodes are smaller than a given threshold value.
The total energy always remains the same, but spreads throughout the network.
\subsubsection{Link-Based Exposure Features}
After a propagation method, such as PR or SPA, has been applied to a network, each node possesses an exposure score that can be viewed as the relative ranking of the node compared to the rest of the network. 
The score can be used as a feature directly or by determining a cut-off value. 
Nodes with an exposure score lower than the cut-off are defined as low-risk nodes and those with an exposure score above the cut-off as high-risk nodes \citep{van2016gotcha}.
Then, based on this re-labelling of the network, new link-based features can be extracted.
This is demonstrated in Figure \ref{fig:propagation}.

\subsection{The Expected Maximum Profit Measure}\label{subsec:EMP}
Model selection highly depends on how the performance is measured.
Traditional measures for credit scoring models include AUC, Gini coefficient and the KS statistic that either assess the discriminative ability of the models or the correctness of the categorical predictions \citep{lessmann2015benchmarking}.
The recently proposed Expected Maximum Profit (EMP) measure has an advantage over these traditional measures because it considers the expected losses and operational income generated by the loan, and is tailored towards the business goal of credit scoring \citep{verbraken2014development}.
Most importantly, when applied to credit scoring models it facilitates computing the models' value, the fifth V of Big Data. 
The measure is based on the expected maximum profit measure, originally developed for customer churn prediction \citep{verbraken2013novel}, and is expressed for credit scoring by
\[
EMP=\int_{b_0}\int_{c_1} P(T(\Theta);b_0,c_1,c^{\ast})\cdot h(b_0,c_1)dc_1 db_0
\]
where 
\[
P(t;b_0,c_1,c^{\ast})=(b_0-c^{\ast})\pi_0 F_0(t)-(c_1+c^{\ast})\pi_1 F_1(t)
\]
is the average classification profit per borrower given the prior probabilities of being a defaulter (non-defaulter), $\pi_0$ ($\pi_1$), and the cumulative density functions of defaulters (non-defaulters), $F_0(s)$ ($F_1(s)$).
Furthermore, $b_0$ is the benefit of correctly identifying a defaulter, $c_1$ the cost of incorrectly classifying a non-defaulter as a defaulter, $c^{\ast}$ the cost of the action, $\Theta=\frac{c_1+c^{\ast}}{b_0-c^{\ast}}$ the cost/benefit ratio and $h(b_0,c_1)$  the joint probability density function of the classification costs \citep{verbraken2014development}.
The maximum profit is achieved by optimizing the cut-off dependent average classification profit where the optimal cut-off value is
\[
T=argmax_{\forall T}P(t;b_0,c_1,c^{\ast}).
\]
As a result, the measure clearly defines an optimal fraction, expressed as
\[
\bar{\eta}_{EMP}=\int_{b_0}\int_{c_1} [\pi_0 F_0(T(\Theta))+\pi_1 F_1(T(\Theta))]\cdot h(b_0,c_1)dc_1 db_0,
\]
representing the fraction of applications that should be rejected to receive maximum profit.
\citet{verbraken2013novel} showed that the EMP corresponds to integrating over the range of the ROC curve that would be considered in a real application, discarding the segment that has a very high, unreasonable cost, and that it is an upper bound of the profit a company could achieve by applying the respective classifier.

When deriving the parameters $b_0 ,c_1$ and $c^{\ast}$ and the probability distribution $h(c_1,b_0)$,  \citet{verbraken2014development} rely on the profit framework discussed in \citet{bravo2013granting}.
Thus, $b_0$ is specified as the fraction of the loan amount that is lost after default or
 \begin{equation}\label{eq:lamdba}
b_0=\frac{LGD\cdot EAD}{A} =: \lambda, 
 \end{equation}
where $LGD$ is the loss given default, $EAD$ is the exposure at default and $A$ the loan amount. 
Furthermore, $c_1$ equals the return on investment ($ROI$) of the loan and $c^{\ast}=0$ since rejecting a customer does not generate any costs.
It only remains to determine $h(b_0,c_1)$ where $ROI(c_1)$ is assumed to be constant but $\lambda(b_0)$ needs to be estimated for each dataset because it is more uncertain with a multitude of possible distributions.

\subsubsection{Model Profit}\label{subsubsec:profit}
The EMP fraction can subsequently be used to compute the profit of a given model.  
First, it is translated into a cut-off value, which depends on the number of instances in the test set.
The instances are labelled as defaulters or non-defaulters depending on whether their predicted score is higher or lower than the cut-off.
Then for each customer in the test set, the confusion matrix in Table \ref{T:confusion} is used to compute the loss or gain produced by the customer.
The model profit is finally computed by aggregating the profit of all customers.
\begin{table}
\centering
\caption{Confusion matrix for computing model profit \label{T:confusion}}
\scalebox{0.85}{\begin{tabular}{llcc}
\hline
&&\multicolumn{2}{c}{Predicted class}\\
&&Non-default&Default\\ \hline
\multirow{2}{*}{Actual class }&Non-default&$ROI\cdot A$&$-ROI\cdot A$\\
&Default&$-LGD\cdot EAD$&0\\ \hline
\end{tabular}}
{}
\end{table}

\subsubsection{Feature Importance in Terms of Profit}\label{subsubsec:feature}
When a credit scoring model is built using the random forest algorithm, its properties can be used to measure the profit impact of each feature in the model.
Assuming a random forest model $RF$ was built using $N$ trees $(T_i)_{i=1}^N$ and $M$ features $(F_j)_{j=1}^M$, the feature importance in terms of profit can be computed in the following way.  
\begin{enumerate}
\item
Apply the random forest model $RF$ to the test set and extract class predictions for each tree $T_i\in RF$.
\item
For each tree $T_i$ compute the profit $P(T_i)$ using the confusion matrix in Table \ref{T:confusion}.
\item
For each feature $F_j$ in the test set, compute the mean decrease in profit.
This is defined as the difference between the average profit of trees where $F_j\in T_i$ and the average profit of trees where $F_j\notin T_i$, given by the equation
\begin{equation*}
P(F_j)=\frac{\sum\limits_{i,F_j\in T_i}P(T_i)}{|\{T_i:F_j\in T_i\}|}-\frac{\sum\limits_{i,F_j\notin T_i}P(T_i)}{|\{T_i:F_j\notin T_i\}|}
\end{equation*}
where $F_j\in T_i$ means that feature $F_j$ is in tree $T_i$.
\item
Sort $P(F_j)$: the features with the highest values are those with the greatest mean decrease in profit.
\end{enumerate}
The result of this method is a ranking of the features in terms of the importance with respect to profit.

\section{Experimental Design}\label{sec:experimentaldesign}
\subsection{Data Description}\label{subsec:datadescr}
The data used in this study originates from a telecommunications operator and a commercial bank that both operate in the same country.
The datasets are anonymized, and do not contain personal information such as the name and address of customers. 
The telco data contains five consecutive months of CDR data of almost 90 million unique cell phone numbers as described in Table 1.
The data from the bank includes over two million customers and it consists of three parts, namely sociodemographic information, such as age, marital status and postcode; debit account activity, including timing and amount of payments; and credit card activity.
Both sociodemographic and debit account activity span three months and conform the historic part of the dataset.
For the credit card activity, there is information about when the cards were issued, the total credit limit, monthly values of how much of the credit remains and how often the customers have failed to repay their debt until twelve months after receiving the card.
The credit card transactions serve as the key input to the credit scoring application because the data provides information about monthly payment arrears. 
This is used to predict the creditworthiness of the customers.
The knowledge about the credit limit and remaining credit on the cards also allows the computation of the EMP.

\subsection{Experimental Setup}
\begin{figure}
\centering
{\includegraphics[scale=0.3]{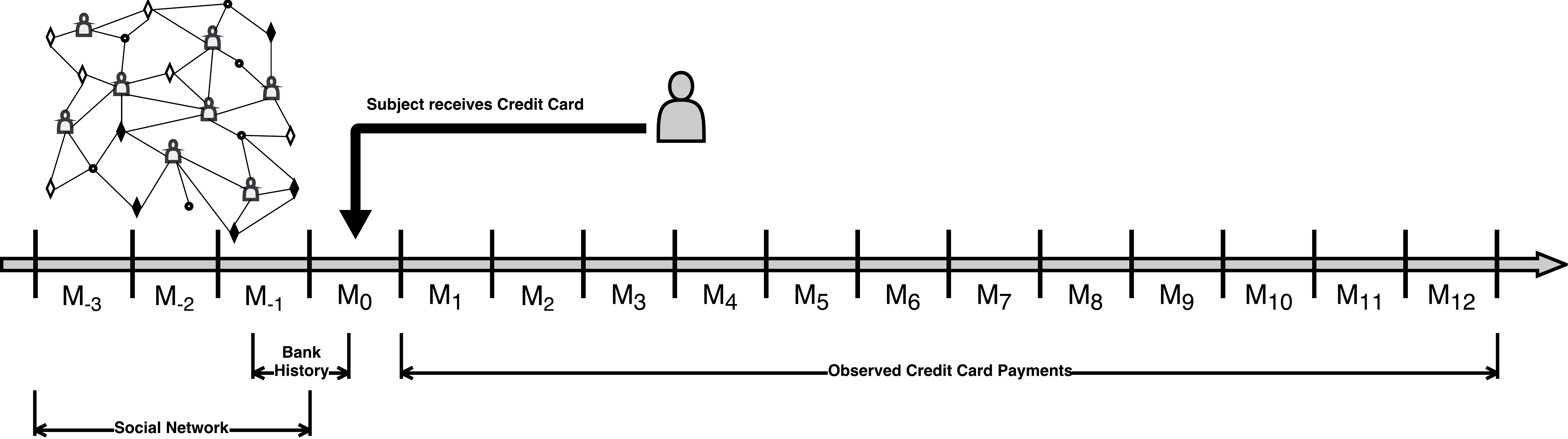}}
\caption{Experimental setup for one timeframe.\label{fig:experimentalSetup}}
\end{figure}

The credit scoring models are built for customers who received a credit card within a three-month period in 2015 and they are referred to as subjects.
An overview of the experimental setup can be seen in Figure \ref{fig:experimentalSetup}.
The credit card data contains information which enables the labeling of the subjects as defaulters or non-defaulters by counting how many late payments they have in the year after signing up for the card.
As previously noted, the Basel definition is used where having three or more late payments implies default.
The label or target vector is denoted by $y_{Default}$. 

To create the bank component of the dataset, both the sociodemographic and debit account data is used.
More precisely, sociodemographic features such as age, marital status and residency as reported at the time of the credit card application are extracted.
Furthermore, debit account activity in the month prior to receiving the credit card is considered and used to extract features representing spending behavior, as can be seen in Table \ref{T:features}.
Based on \citet{singh2015money}, two types of temporal-behavioral features that have been shown to correlate with financial well-being and consumption are included. 
The first one, diversity, measures how customers spread their transactions over various bins, represented by the days of the week in this case.
For each customer $i$ and each bin $j$, the fraction of transactions $p_{ij}$ that fall within bin $j$ is computed.
The temporal diversity of customer $i$ is then defined as the normalized entropy of all transactions counted in all seven bins with $M$ being the number of non-empty bins, or
\[
D_i=\frac{-\sum_{j=1}^7 p_{ij}\log p_{ij}}{logM}.
\]
In addition, the loyalty of a customer is defined as
\[
L_i=\frac{f_i}{\sum_{j=1}^7 p_{ij}}
\]
where $f_i$ is the fraction of all transactions of customer $i$ that happen in their $k$ most frequently used bins.
In this case, loyalty characterizes the percentage of transactions that take place during a customer's three most active days.
The collection of both sociodemographic and debit account features is called `sociodemographic' features and denoted with $x_{SD}$.

\begin{figure}
\centering
{\includegraphics[scale=0.27]{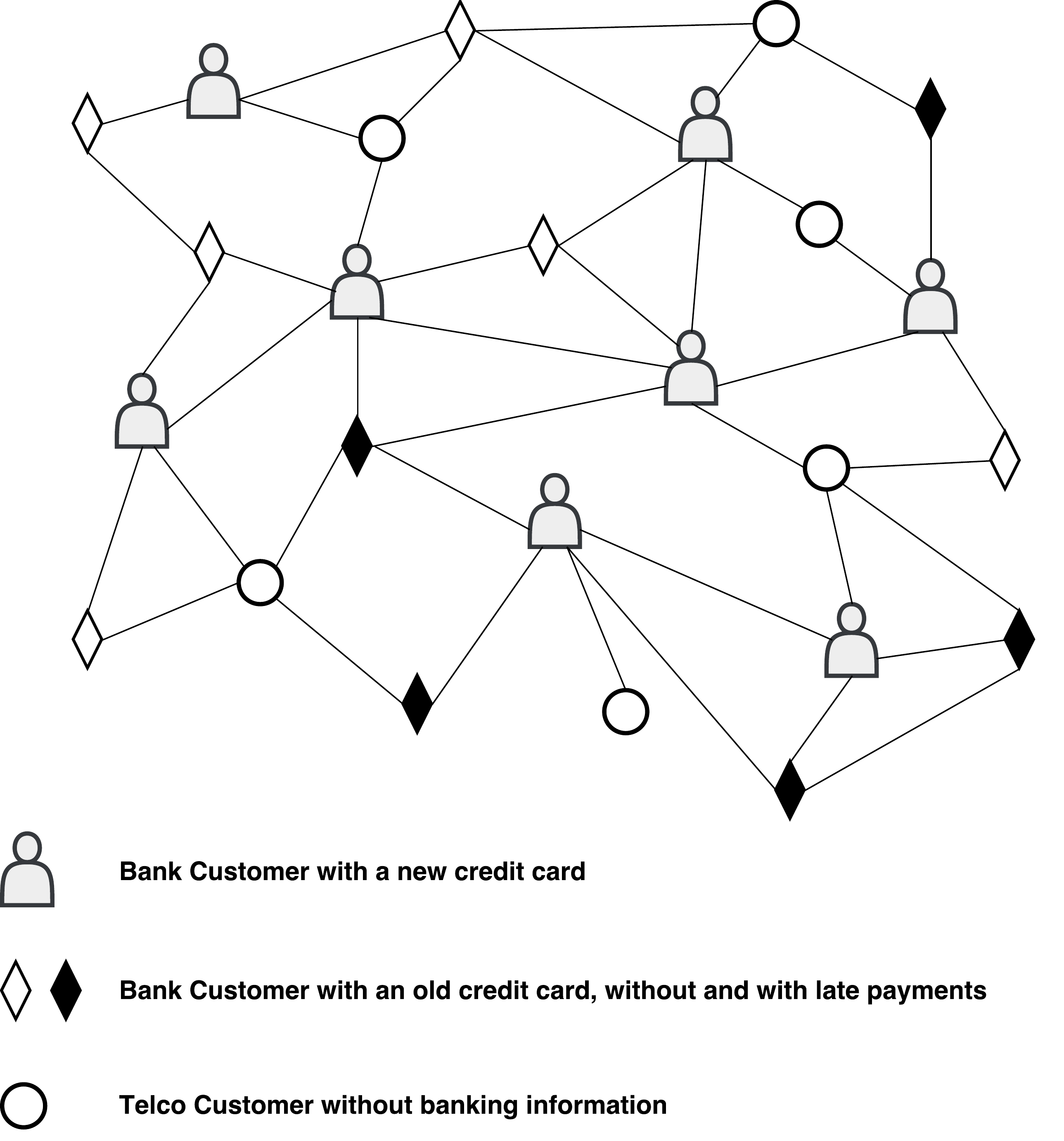}}
\caption{The figure demonstrates the various types of people that are present in the call network. \label{fig:TypesOfPeople}}
\end{figure}
The telco data is used as the key input for the social network part of the analysis.
As mentioned before, the subjects received their credit cards within a period of three months and the subjects are considered in each month separately, which results in three timeframes $t_1$, $t_2$ and $t_3$.
To build a call network for each timeframe, the CDR of three whole months prior to the card acquisition month is aggregated and people that have shared a phone call during this period are linked together after discarding any phone calls lasting less than five seconds.
Thus, there are three call networks spanning three months each.
Each network consists of all subjects that received a credit card in the month succeeding the last month in the network, everyone they shared a phone call with and all phone calls between everyone in the network.  
In addition to the subjects, there are also other types of people in the network, as Figure \ref{fig:TypesOfPeople} shows.
The people-shaped entities are the subjects, whereas the diamond-shaped entities denote other bank customers (i.e., people who did not receive a credit card during the three months).
They may, however, already possess a card, and those that are known to have had payments arrears are colored black. 
These are the delinquent customers in the network, as described in section \ref{subsec:callNetworks}. 
Bank customers without payment arrears are colored white.
The circular entities in the network are people who are customers of the telco but not of the bank.  

For all subjects in each of the three timeframes, four types of network features, both direct and indirect, are extracted. 
First, features representing the calling behavior of the subjects. 
Thus, the number and duration of incoming, outgoing and undirected phone calls taking place during the day and night and on different days of the week are computed.
These features are denoted by $x_{CB}$.
As described in subsection \ref{subsec:callNetworks}, information about delinquent customers in the network--the black diamonds--is used and they are labeled with respect to three distinct criteria: having one or more late payments, having two or more late payments and, having three or more late payments.
This gives the opportunity to distinguish the severity of their financial situation in relation to the influence they spread.
These three label vectors serve as the information source $z$ and active nodes $V^A$ when applying PR and SPA, respectively.  
Based on these labellings the extraction of link-based features, computation of PR and SPA exposure scores together with link-based exposure features as described in subsection \ref{subsec:callNetworks} is performed. 
To construct the weight matrix $W$, edges in all networks are weighted by the number of phone calls and both incoming and outgoing edges as well as undirected networks are considered.
The parameters in the propagation algorithms are set to the default values $\alpha=0.85$ (for PR) and $d=0.85$ (for SPA), based on exploration of the data which showed robust results.
For the link-based exposure features, the cut-off point is defined as the minimum exposure score of the delinquent customers with at least three late payments, since having at least three late payments defines default.
All the link-based features are viewed as one group of features denoted by $x_{LB}$. 
Finally, the feature groups $x_{PR}$ and $x_{SPA}$ are respectively composed of the exposure scores of PR and SPA together with the corresponding link-based exposure features.

The result of the featurization process is a dataset of the form
\[
x=\{x_{SD},x_{CB},x_{LB},x_{PR},x_{SPA}\}, \quad y=\{y_{Default}\}
\]
Table \ref{T:features} describes some of these features.
After combining the two data sources, extracting all the features described above and cleaning up the dataset, 22,000 observations remain and over 300 features.
The fraction of defaulters is $0.0449$ or just under $5\%$ default rate.

With the datasets featurized, credit scoring models are built using binary classifiers with a $70\% / 30\%$ split into training and test set.
Before building the models, highly correlated variables are removed and undersampling of the training set conducted to reduce class imbalance, as is common when applying analytics techniques \citep{baesens2014analytics}.
Final model performance is evaluated using the test set.
The binary classifiers logistic regression, decision trees and random forests are used for the empirical analysis.  
Logistic regression is the industry standard for building credit scoring models \citep{scheule2016credit}. 
Decision trees are included since they are more powerful than logistic regression, while at the same time guaranteeing interpretability of the model.
They are implemented using recursive partitioning with ten-fold cross validation on the training set to tune and prune the trees.
Both the logistic regression and decision tree models are compared against random forests which are an ensemble method that constructs multiple decision trees that jointly decide upon the credit score.  
Random forests are considered to be a very powerful, black-box analytical modeling technique. 
As a result of parameter tuning, 500 trees were used to build each forest.

\begin{table}
\caption{Descriptions of some of the features that were extracted from the data sources.  In the table IN, OUT and UD stand for networks with incoming, outgoing and undirected edges, respectively.  The number in the brackets (x) indicates how delinquent customers were defnined with respect to the number of payment arrears in each case. \label{T:features}} 
\scalebox{0.6}{
\begin{tabular}{m{1.5cm}lm{1.2cm}rp{18cm}}
Feature Group&Notation&Number&Feature&Description\\\hline
\multirow{8}{1.5cm}{Socio demo graphic}&\multirow{8}{*}{SD}&\multirow{8}{*}{35}&Age&Current age of the customer\\
&&&Amount Spent&Total amount spent in the month before receiving the credit card \\
&&&Mean Spent p. Day& Average amount spent per day during the month before receiving the credit card \\
&&&Diversity-NE Value&Diversity of value spent over non-empty bins during the month prior to receiving the credit card\\
&&&Diversity-ALL Number&Diversity of number of purchases over all seven bins during the month prior to receiving the credit card\\
&&&Loyalty-Number&Loyalty of number of purchases in the top three bins during the month prior to receiving a credit card\\ \hline
\multirow{3}{2cm}{Calling Behavior}&\multirow{3}{*}{CB}&\multirow{3}{*}{72}&Count IN&Total number of phone calls received during the three months of the social network \\
&&&Weekend Duration OUT&Aggregated duration of all phone calls made on weekends during the three months of the social network   \\
&&&Tuesday Duration UD&Aggregated duration of all phone calls made and received on Tuesdays during the three months of the social network \\\hline
\multirow{8}{1cm}{Link-Based}&\multirow{8}{*}{LB}&\multirow{8}{*}{36}&Binary (0) IN&Binary indicator of having neighbors with no late payments, in a network with incoming edges\\
&&&Binary (1) OUT&Binary indicator of having neighbors with one late payment, in a network with outgoing edges\\
&&&Binary (2) UD&Binary indicator of having neighbors with two late payments, in a network with undirected edges\\
&&&Binary (3) UD &Binary indicator of having neighbors with three late payments, in a network with undirected edges\\
&&&Count (0) IN&Number of neighbors with no late payments, in a network with incoming edges \\
&&&Count (1) OUT&Number of neighbors with one late payment, in a network with outgoing edges\\
&&&Count (2) OUT&Number of neighbors with two late payments, in a network with outgoing edges\\
&&&Count (3) UD&Number of neighbors with three late payments, in a network with undirected edges\\\hline
\multirow{6}{1.5cm}{Persona-lized PageRank}&\multirow{6}{*}{PR}&\multirow{6}{*}{54}&Exposure (1) IN&Exposure score after applying PR on a network with incoming edges and delinquent customers with one or more late payments.\\
&&&Exposure (2) OUT&Exposure score after applying PR on a network with outgoing edges and delinquent customers with two or more late payments.\\
&&&Exposure (3) UD&Exposure score after applying PR on a network with undirected edges and delinquent customers with three or more late payments.\\
&&&Binary High Risk (1) IN&Binary indicator of having neighbors with high exposure scores  after applying PR on a network with incoming edges and delinquent customers with one or more late payments. \\
&&&Binary High Risk (2) OUT&Binary indicator of having neighbors with high exposure scores  after applying PR on a network with outgoing edges and delinquent customers with two or more late payments. \\
&&&Count High Risk (3) IN&Number of neighbors with high exposure scores after applying PR on a network with incoming edges and delinquent customers with three or more late payments.\\\hline
\multirow{6}{2cm}{Spreading Activation}&\multirow{6}{*}{SPA}&\multirow{6}{*}{54}&\\
&&&Exposure (1) IN&Exposure score after applying SPA on a network with incoming edges and delinquent customers with one or more late payments.\\
&&&Exposure (2) OUT&Exposure score after applying SPA on a network with outgoing edges and delinquent customers with two or more late payments.\\
&&&Exposure (3) UD&Exposure score after applying SPA on a network with undirected edges and delinquent customers with three or more late payments.\\
&&&Binary High Risk (1) IN&Binary indicator of having neighbors with high exposure scores  after applying SPA on a network with incoming edges and delinquent customers with one or more late payments. \\
&&&Count High Risk (1) UD&Number of neighbors with high exposure scores after applying SPA on a network with undirected edges and delinquent customers with one or more late payments. \\
&&&Count High Risk (3) IN&Number of neighbors with high exposure scores after applying SPA on a network with incoming edges and delinquent customers with three or more late payments.\\\hline
\end{tabular}}
\end{table}

\section{Results}\label{sec:results}

The results are organized in three parts starting with empirical tests to establish the networks' relational dependency.
Subsequently, the results of the proposed methodology are detailed, first in terms of statistical performance and then in terms of economic performance.

\subsection{Homophily amongst Defaulters} \label{sec:hom}
A network is homophilic if nodes with a certain label are to a larger extent connected to other nodes with the same label.
In the default networks, homophily is present if the fraction of edges between defaulters and non-defaulters is significantly smaller than the expected fraction of such edges in the network.
A one-tailed proportion test with a normal approximation for homophily amongst defaulters resulted in a p-value of less than 0.0001, which means that there is evidence of homophily \citep{baesens2015fraud}.
Furthermore, homophily in networks can also be measured with dyadicity and heterophilicity, that is, the connectedness between nodes with the same label and of different labels, respectively, compared to what is expected in a random network \citep{baesens2015fraud}. 
The networks analyzed here, have a dyadicity amongst defaulters of 0.8689 while the heterophilicity is 0.8137. 
This means that the networks are not dyadic, as defaulters are not more connected amongst themselves, but they are heterophilic, i.e., there are less connections between defaulters and non-defaulters.
Based on these results, there is foundation for applying social network analytic techniques to predict default in the call networks.

\subsection{Statistical Model Performance}
\begin{table}
\centering
\caption{Statistical Model Performance (AUC). \label{T:modelperformance}}
\scalebox{0.85}{\begin{tabular}{llccc}
\hline
\multicolumn{2}{c}{Model}&\multicolumn{3}{c}{Classifier}\\ 
Model ID&Feature Groups&Logistic Regression&Decision Trees&Random Forest\\ \hline
A&SD&0.5869&0.7004&0.8993\\
B&CB&0.5351&0.7043 &0.8700\\
C&LB&0.5485&0.7429 &0.7697\\
D&PR&0.5163&0.7611 &0.8339\\
E&SPA&0.5281&0.7188 &0.8063\\ 
F&SD,CB&0.6115&0.7127&0.9227\\
G&CB,LB,PR,SPA&0.5182&0.7307&0.9154\\ 
H&SD,CB,LB,PR,SPA&0.6121&0.7263&0.9224\\ \hline
\end{tabular}}
\end{table}

Credit scoring models are built with the features in each feature group separately, as well as three models with a combination of feature groups, as seen in Table \ref{T:modelperformance}.
The first five models $A$, $B$, $C$, $D$ and $E$ study the main effects of each feature group. 
Model $F$ combines the sociodemographic features with the calling behavior features, model $G$ includes all feature groups except the sociodemographic features and in model $H$ we consider all feature groups.
Other combinations of feature groups were tried, but they did not provide more significant results than the ones shown.
As is common practice in credit scoring, statistical model performance is measured by the area under the receiver operating curve (AUC).
The AUC summarizes the trade-off between model sensitivity and specificity in a single number between 0 and 1 with higher values meaning better performance.

From Table \ref{T:modelperformance}, it is clear that the performance with respect to the three classifiers varies substantially.
Overall, the logistic regression models perform the worst, of which models including sociodemographic features (models $A$, $F$, $H$) perform best.
Logistic regression models do not yield a better performance when using network-related features. 
This hints at a non-linear behavior that cannot be properly captured by a generalized linear model.

\begin{figure}%
\centering
{    \subfloat[$95\%$ Confidence]{{\includegraphics[clip, trim=5cm 18.5cm 12.5cm 4cm, width=0.35\textwidth]{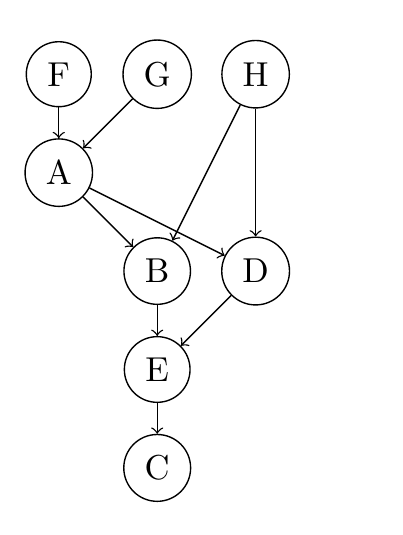} }}%
    \qquad
    \subfloat[$99\%$ Confidence]{{\includegraphics[clip, trim=5cm 18.5cm 12.5cm 4cm, width=0.35\textwidth]{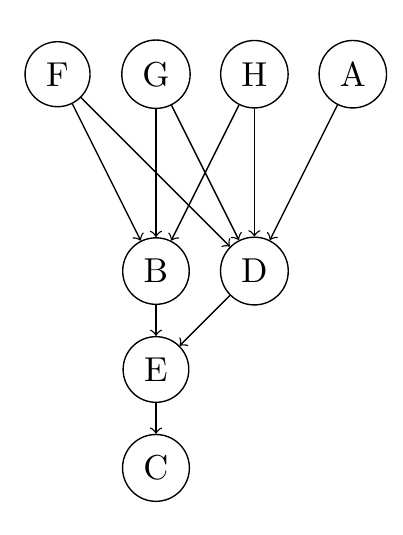} }}}%
 \caption{Domination graphs. 
 \label{fig:domination}}
\end{figure}

The random forests produce the best-performing models and the remaining discussion will therefore be focus on them.
First, the test of \citet{delong1988comparing} is applied  to the receiver operating curves (ROC) of each pair of random forest models to compare their performance.  
The results can be seen in the domination graphs in Figure \ref{fig:domination}.
The best performing models are at the top and models that perform worse are lower down. 
The arrows indicate a significant improvement in statistical performance at $95\%$ and $99\%$ confidence level on the left and right, respectively.
The figure on the left demonstrates that there is not a significant difference in the performance of the three models with a combination of features ($F$, $G$, $H$), but models with only one type of features ($A$, $B$, $C$, $D$, $E$) perform significantly worse with the link-based features ($C$) performing worst overall.
\begin{figure}
\centering
{\includegraphics[scale=0.8]{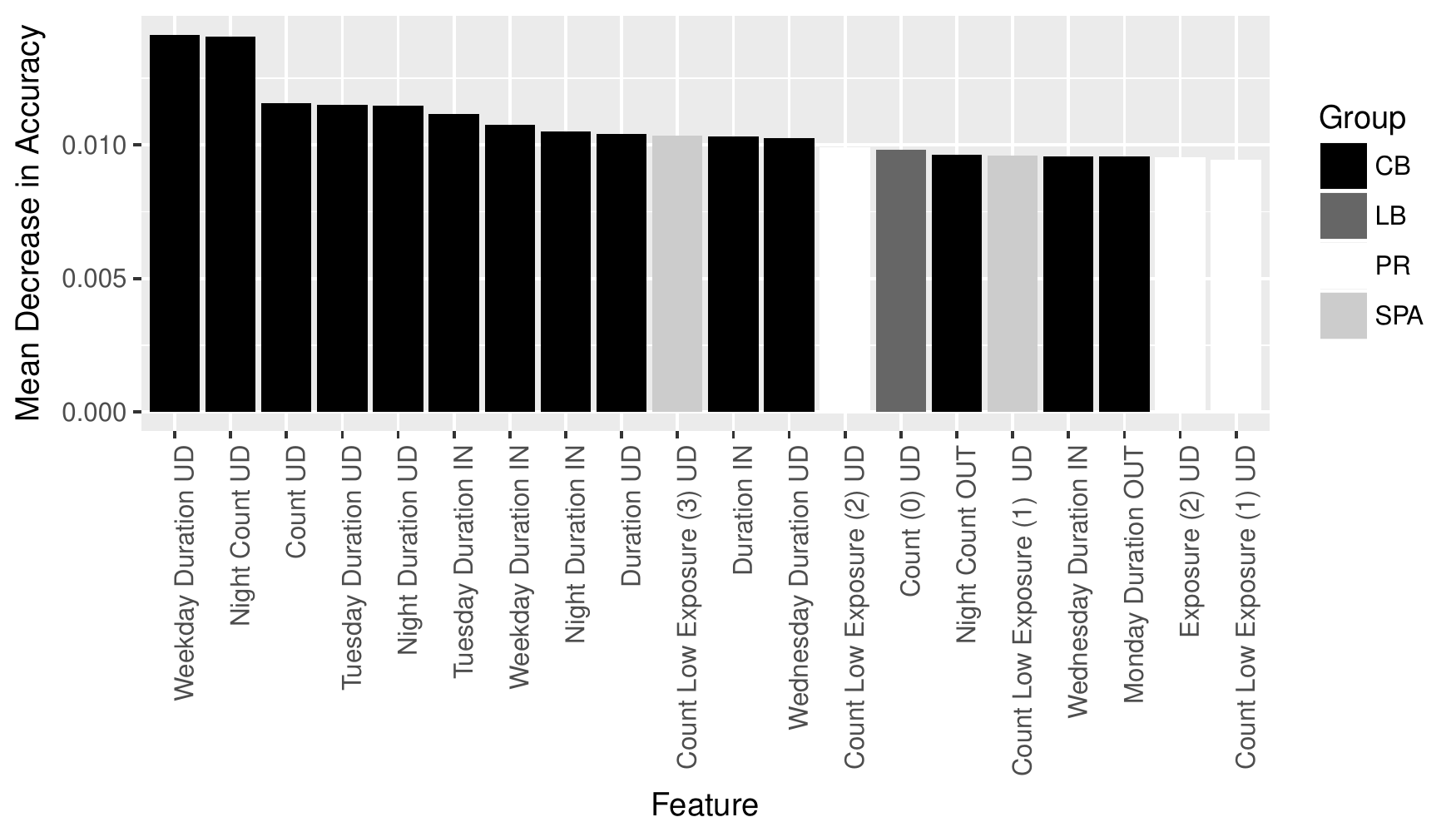}}
\caption{Feature importance: Mean decrease in accuracy.
\label{Fig:varimpAcc}}
\end{figure}
Secondly, the importance of the features in model $H$ is explored to determine their ability to predict default and rank the usefulness of the features. 
This is displayed in Figure \ref{Fig:varimpAcc} for the mean decrease in accuracy for the 20 most important variables.
The mean decrease in accuracy of a particular feature measures how much the accuracy of the resulting model decreases when that feature is left out of the model, and as a result, gives a score of how important it is in the model.
Figure \ref{Fig:varimpAcc} demonstrates that the calling behavior features are ranked the highest, followed by PR features and SPA features, and a single LB feature.

\subsection{Economic Model Performance}
The previous subsection showed that the statistical performance of more complex credit scoring models with a combination of feature groups is significantly better than models with only one feature group, and even better than that of models with sociodemographic features alone.
Here the economic performance of the models is evaluated and the importance of features in terms of profit by applying the EMP to the random forest models.
\begin{figure}%
\centering
  {  \subfloat[EMP]{{\includegraphics[width=0.45\textwidth]{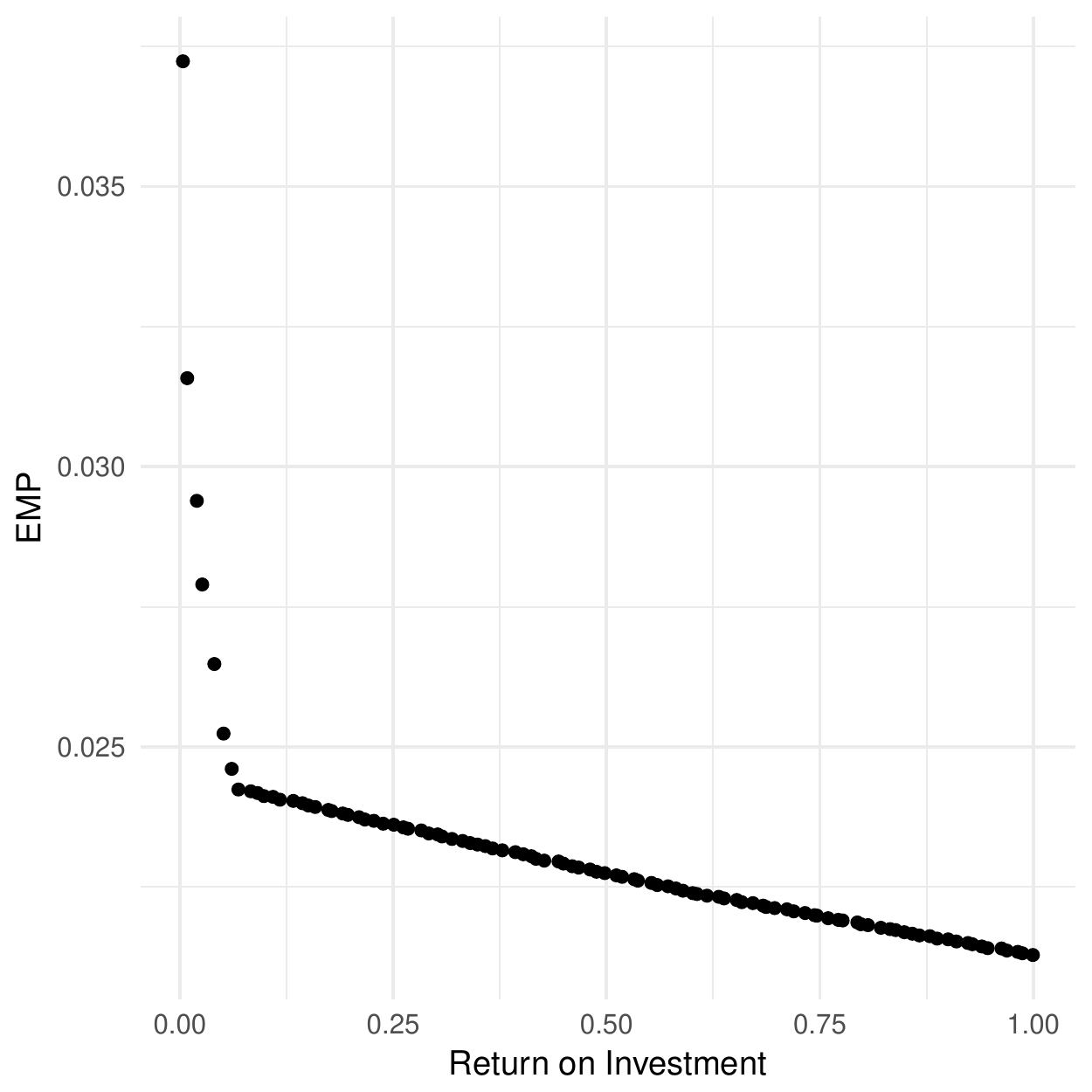} }}%
    \qquad
    \subfloat[EMP fraction]{{\includegraphics[width=0.45\textwidth]{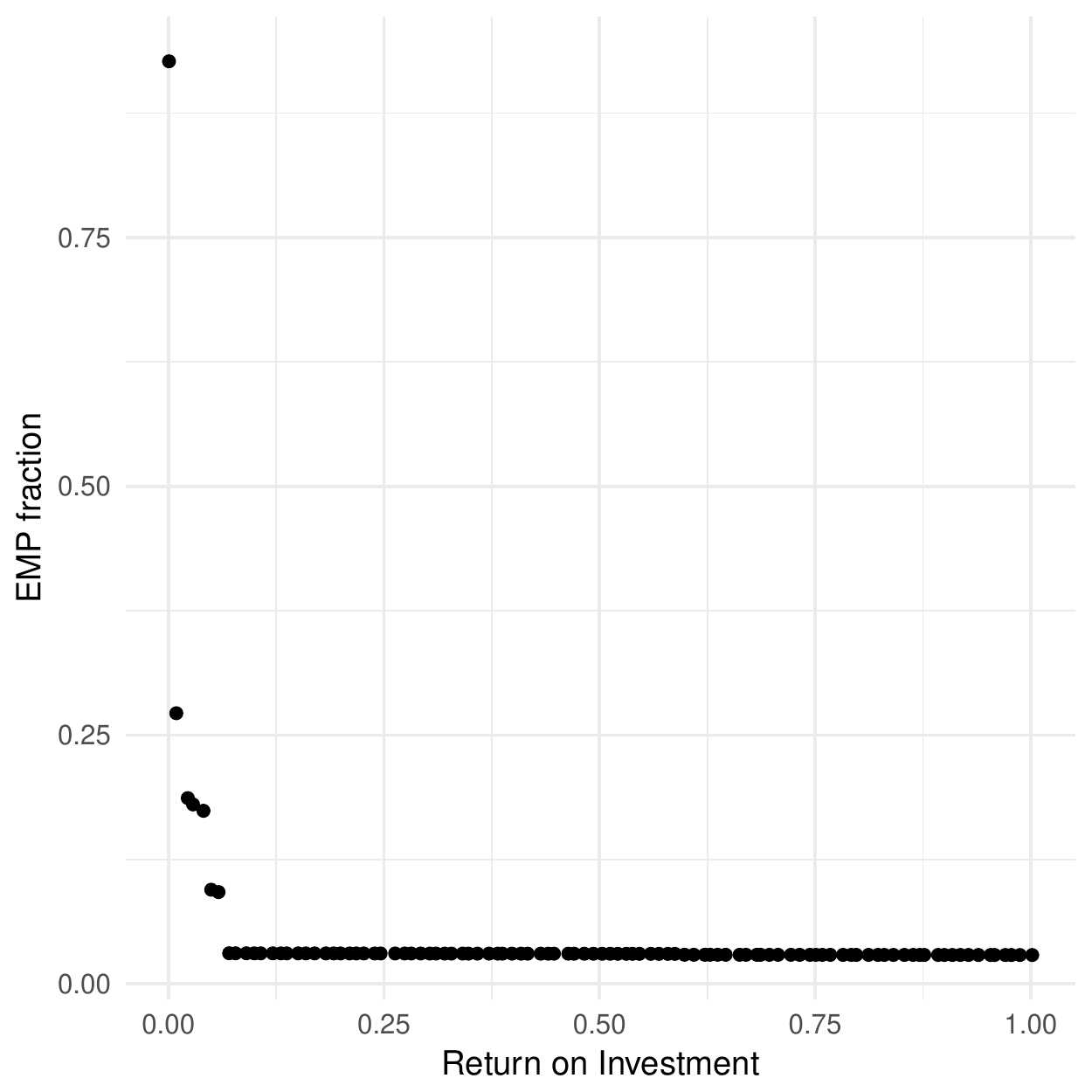} }}}
     \caption{Sensitivity Analysis for ROI. 
     \label{fig:sens2D}}
\end{figure}
For the EMP (see section \ref{subsec:EMP}) various parameters need to be specified.
To compute the benefit of correctly identifying a defaulter, $\lambda$ (see Equation \ref{eq:lamdba}), the credit card limit is used as the principal ($A$) and the drawn amount on the card at the time of default as exposure at default ($EAD$).
The two remaining parameters: loss given default ($LGD$) and return on investment ($ROI$), are domain specific and not obtainable from the data directly.
Therefore, an exploration of their effect on the EMP is provided. 
An analysis of the variation in EMP as a function of $LGD$ shows substantial robustness, which means that the economic performance of the models does not greatly depend on $LGD$.
Considering this, and based on expert judgement, this parameter is set to 0.8.
In contrast, EMP decreases when $ROI$ increases as is evident from Figure \ref{fig:sens2D}, which shows the EMP and its implied cutoff (EMP fraction) as a function of $ROI$ when $LGD$ is set at 0.8.
The value for $ROI$ is determined based on the `elbow' in these figures and set to 0.05.
The inflection point is the point where the $ROI$ becomes the biggest influence (thus the linear behavior) and so it is appropriate to choose a value that balances profits for the rest of the analyses.
\begin{figure}
\centering
{\includegraphics[scale=0.55]{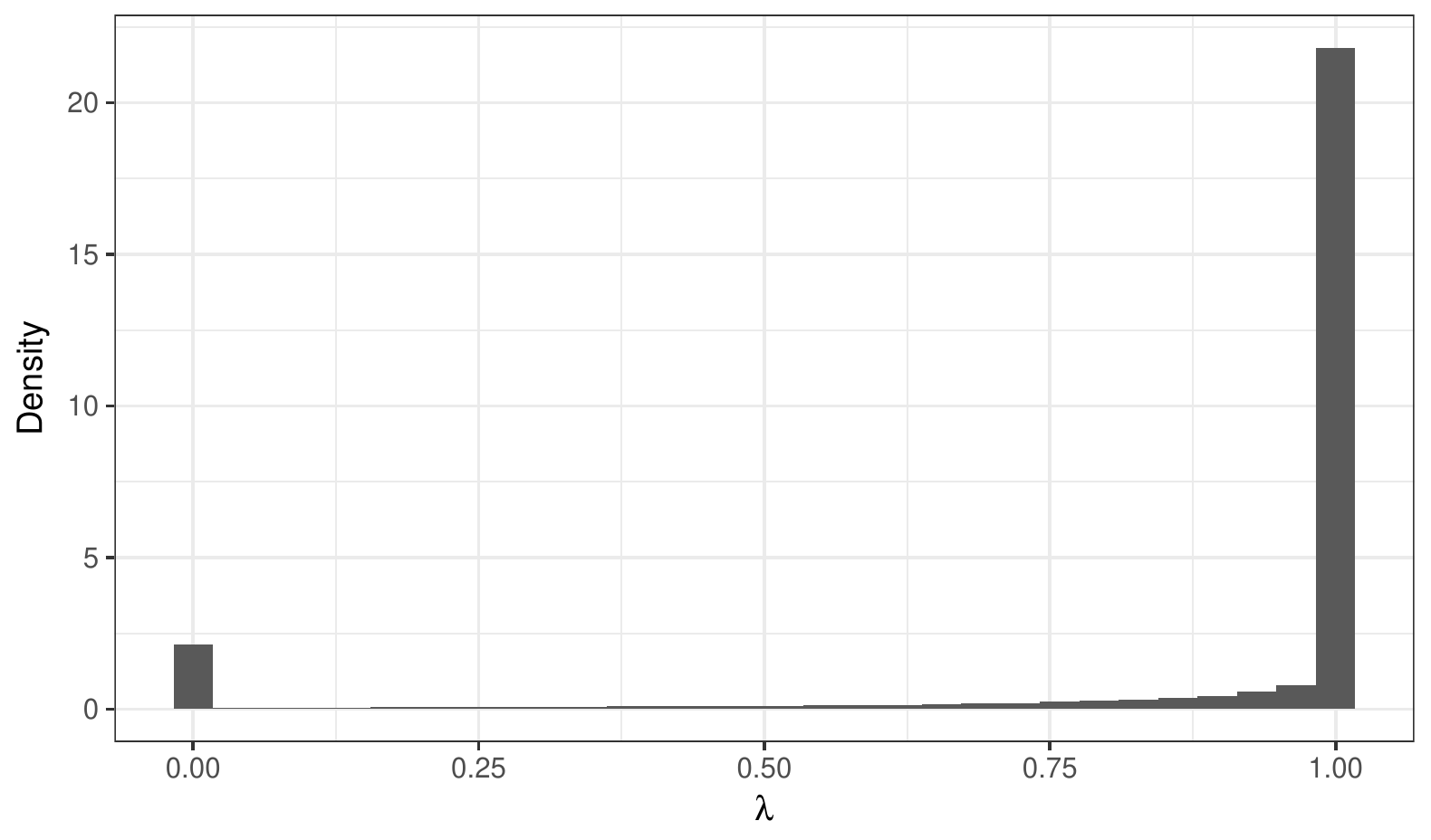}}
\caption{Distribution of $\lambda$.\label{fig:lambda} }
\end{figure}
Subsequently, the distribution of $\lambda$ can be estimated, see Figure \ref{fig:lambda}.
As in \citet{verbraken2014development} there are two peaks in the distribution, one at each end of the unit interval and with the assumption that $\lambda$ follows a uniform distribution in between.
The peak at 0 represents credit card holders who have had payment arrears and have paid back fully, whereas the peak at 1 indicates those that never paid back their debt.
This distribution is used to determine the values for $p_0$ and $p_1$, see \citet{verbraken2014development}.

\begin{figure}%
\centering
  {  \subfloat[EMP and EMP fraction]{{\includegraphics[width=0.45\textwidth]{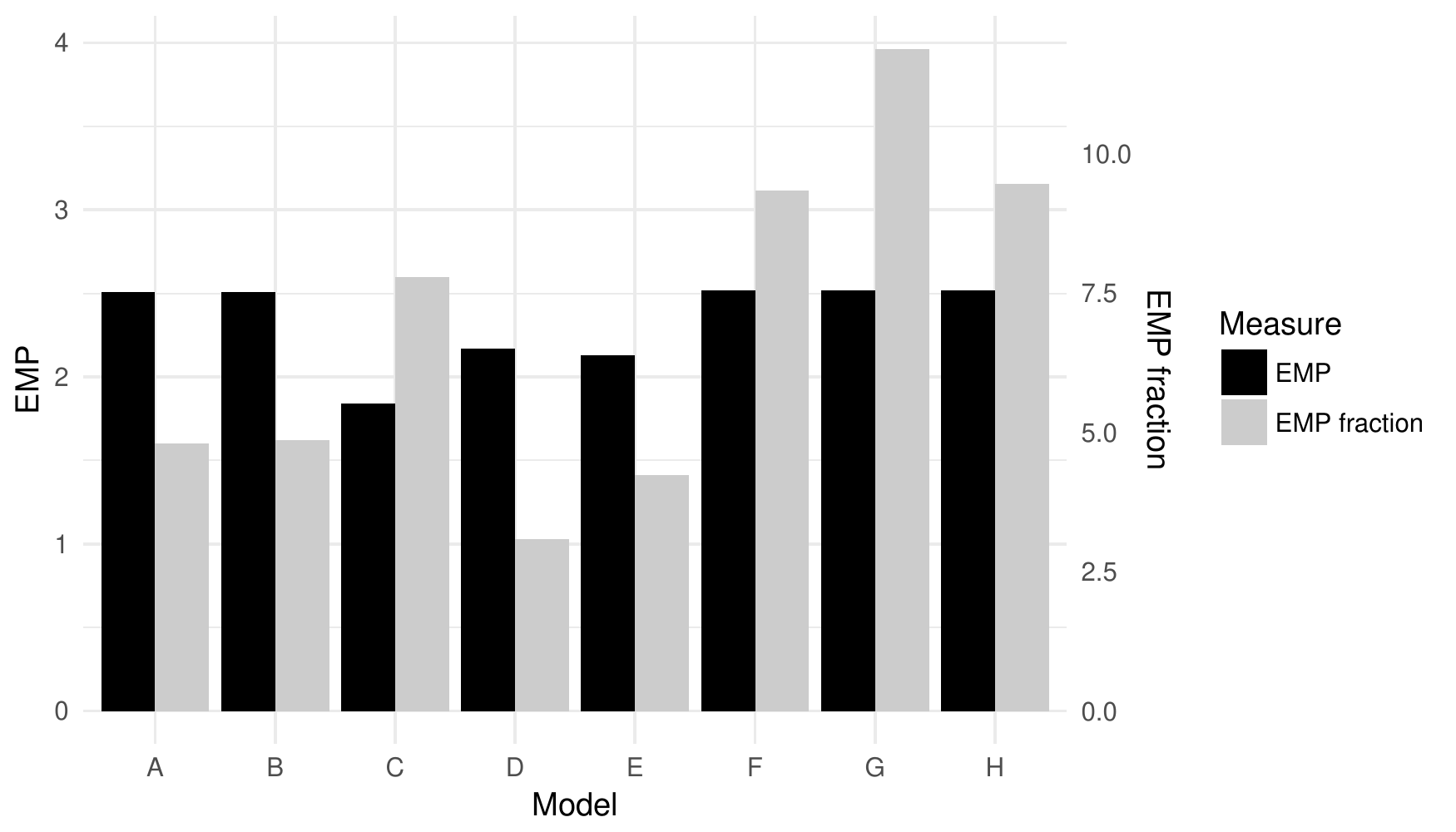} }}%
    \qquad
    \subfloat[Profit]{{\includegraphics[width=0.45\textwidth]{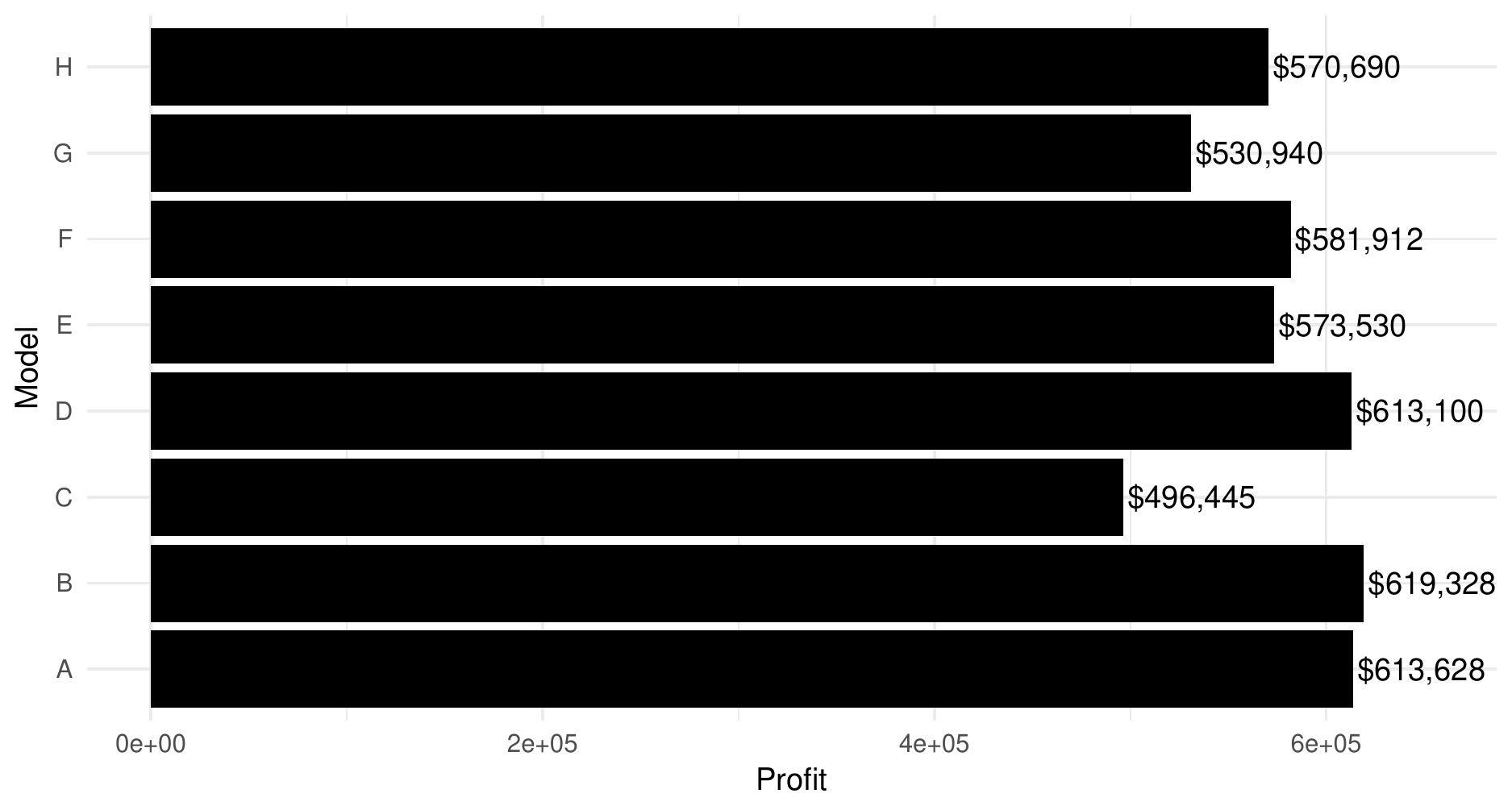} }}}
    \caption{Economic Model Performance. \label{fig:EMP}}
  \end{figure}

With all parameters estimated, the next step is to compute the expected maximum profit, the profit maximizing fraction of rejected loans and the model profit (as described in Section \ref{subsubsec:profit}) for the random forest models in Table \ref{T:modelperformance}.
The results can be seen in Figures \ref{fig:EMP}.
The value for EMP is expressed as a percentage of the total loan amount and measures the incremental profit relative to not building a credit scoring model.  
The ranking of the values for the expected maximum profit is consistent with the ranking of the AUC values in Table \ref{T:modelperformance}, and again models $A$, $F$, $G$ and $H$ are considered best and $C$ the worst.
The EMP fraction values vary, however, and therefore so do the model profits.
The profit maximizing fraction represents the fraction of credit card applications that should be rejected in order to obtain the maximum profit.  
The fact that the fraction for model $G$ is so much higher than the rest with the profit remaining the same, indicates that the model focuses on the most profitable customers.
Regarding the model profits, there is a substantial increase compared to not using a model.
Model $B$ has the highest profit, followed by models $A$ and $D$.
Models $F$ and $H$ also produce decent profits, whereas $G$ does not, at least not when compared to the rest.
Again, model $C$ performs the worst.
Of the models with only one data source, model $B$ (built with calling behavior variables) brings the higher singular results. 
As no history is available for these borrowers, a possible explanation is that their socioeconomic standing can be deduced from their immediate network. 
Note however that this difference is marginal, as model $A$ (sociodemographic variables) follows it. 
Of the combined models, models $F$ (with both sociodemographic variables and calling behavior) and $H$, with all available variables, produce the best results in terms of profits.

\begin{figure}%
\centering
{\includegraphics[width=0.8\textwidth]{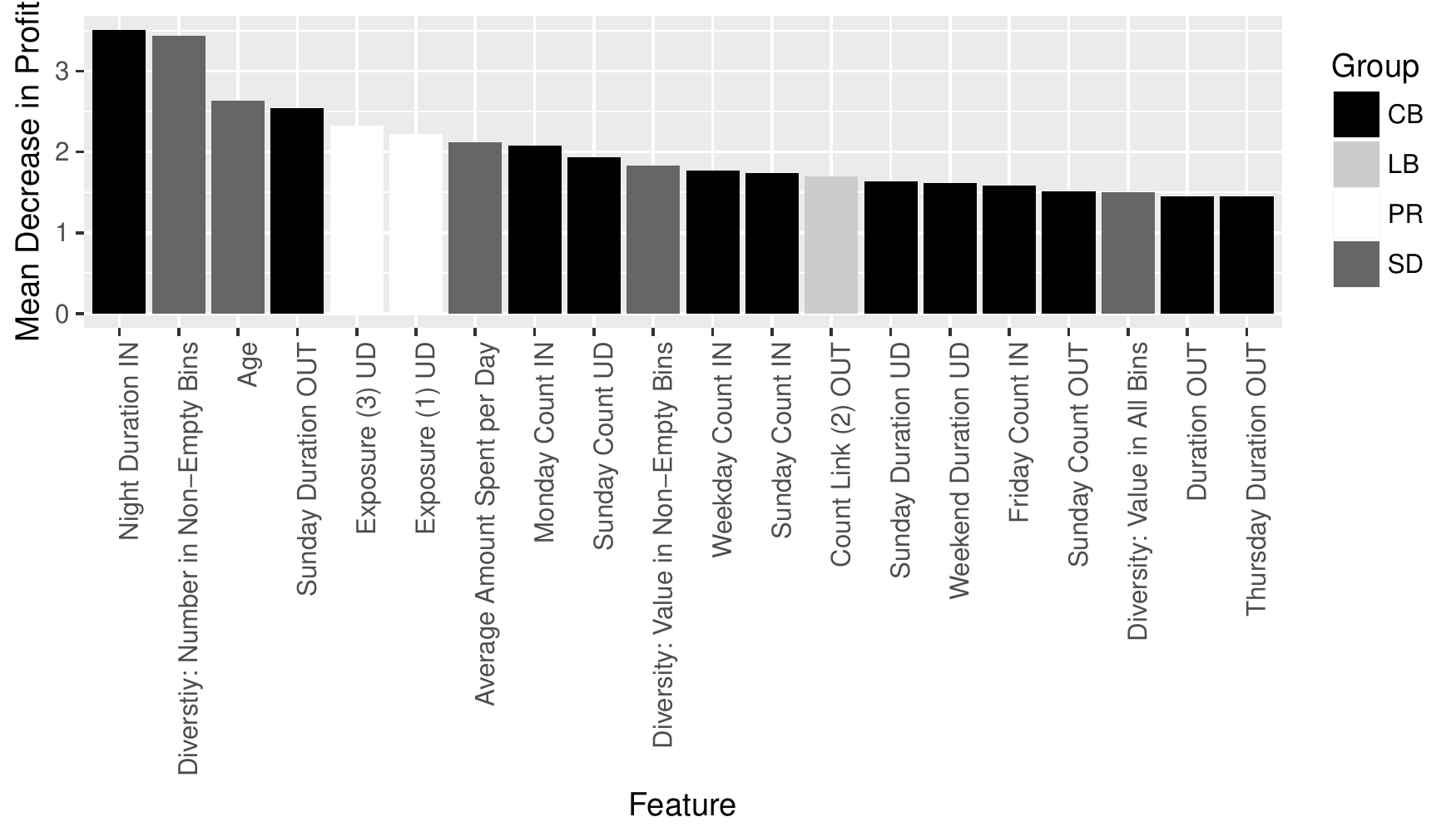} }
\caption{Feature Importance: Mean decrease in Profit.
 \label{fig:varImpEMP} }%
\end{figure}

Figure \ref{fig:varImpEMP} shows the mean decrease in profit for the 20 most important features in model $H$ computed using the technique described Section \ref{subsubsec:feature}. 
This profit perspective shows more variation in groups of features than the statistical one. 
As for mean decrease in accuracy, more than half of the features are calling behavior features, but in contrast to Figure \ref{Fig:varimpAcc}, a quarter of the features are sociodemographic features, which in this case are features that measure consumption.
This is consistent with the result indicating that, of the combined models, model $F$ was associated to a larger profit.
To test for correlation among the ranking of features according to the two measures we computed the Spearman's $\rho$, Kendall's $\tau$ and Goodman and Kruskal's $\gamma$ correlation coefficients.
The resulting values did not indicate a correlation among the rankings.

It is interesting to see that having only network features allows us to discriminate potentially better customers. That would mean that we can look, using the network connections, beyond simple socioeconomic and sociodemographic traits, and actually profile more profitable customers. 
In the models, network features help discriminate different customers, who cannot be captured by common features, and it happens that these customers bring a lot of profit.

\section{Discussion}\label{sec:discuss}
Based on these results, the three research questions in section \ref{secIntroduction} can be addressed.

The first research question Q1 assesses the value that call data adds to credit scoring models in term of AUC and  profit.
For statistical performance, models including all features performed best, with the AUC value increasing by 0.023 points in the best model when compared to the sociodemographic model $A$.
The economic performance of the models in terms of EMP, EMP fraction and profit can be seen in Figure \ref{fig:EMP}.
The model with the highest profit is model $B$ and it is slightly better than the traditional model $A$.
Models with a combination of feature groups ($F$ ,$G$ and $H$) produce lower profit but their EMP values are the highest.
The reason for the lower profit is the high EMP fraction, which indicates that these models are more conservative and exclude a higher proportion of the defaulters.
These results indicate that the CDR data complements the conventional data and there is added value when including the CDR data in credit scoring models and even when used without the traditional features. 

The results also provide an answer to the second research question Q2: Can call data replace traditional data used for credit scoring?
In terms of both statistical and economic performance, the results indicate that the predictive power of call data is just as good or might be even better than traditional data for these borrowers.
This is clear from the high performance of model $B$.
In addition, the importance of the calling behavior features shows that these are very predictive, much more so than the traditional features.
This result demonstrates the merit of this research. Given the high predictive power of the call data, borrowers without enough bank information can benefit from the approach by giving access to their call records to obtain credit.

Finally, the last research question Q3 about how default behavior propagates in the network, can be addressed.
The results of the homophily test in section \ref{sec:hom} showed a lower fraction of connections amongst defaulters and also between defaulters and non-defaulters. 
This might partially be a consequence of the low number of defaulters in the network overall. 
Furthermore, insights about the propagation of default behavior can be obtained from the importance of the features.
Firstly, for mean decrease in accuracy, see Figure \ref{Fig:varimpAcc}, a few PageRank and Spreading Activation features are important.
These are predominantly `Count Low Exposure' features which represent the number of neighbors with low exposure score.
This indicates that not having a high-risk neighbor is predictive of non-default. 
The PageRank feature `Exposure (2)' is also among the 20 most important features.
It represents the PageRank exposure score when the influence comes from delinquent customers with two or more late payments and indicates a propagation effect of default behavior.
Second, for the mean decrease in profit in Figure \ref{fig:varImpEMP} there are two PageRank exposure scores, based on one and three late payments of delinquent customers. 
From these observations we can say that, in terms of propagation of default influence, Personalized PageRank is more effective than Spreading Activation.
A more thorough analysis of how default propagates is needed to better understand the effect of each of these features.

\section{Impact of Research}\label{sec:impact}
The research findings presented in this paper have possible impact at various levels.
This section identifies three different levels and provides a discussion of the implications of each one.

\subsection{Regulatory Impact}
The Basel Accords model unexpected losses using a Merton single-factor model where the asset value of an obligor depends upon a systematic (e.g., the macroeconomy) and an idiosyncratic (e.g., obligor-specific) risk component \citep{scheule2016credit}.  
Asset correlations are then also factored in to see how default behavior is correlated and, as such, model system risk.  
A key concern relates to the exact values of these asset correlations.  
For corporates, the assets can be quantified by inspecting balance sheets, and various financial models have been introduced to quantify corporate asset correlations. 

For retail exposures (e.g., credit cards, mortgages, installment loans), it becomes considerably more difficult as the assets are less tangible.  
Retail asset correlations have been specified in the Basel Accords using some empirical, but not published, procedure reflecting a combination of supervisory judgment and empirical evidence.  
As such, they are fixed at $4\%$ for qualifying revolving exposures (e.g., credit cards) and $15\%$ for mortgages.  
Given their impact on capital calculation, it would be desirable that these asset correlations are sustained by a solid theoretical framework and accompanying empirical validation.  
In this research, we illustrated how default behavior on credit cards propagates in a call network.  
These insights pave the way for additional research aimed at quantifying asset and default correlation for retail exposures in a more sound and solid way.  
This can then lead to better regulatory asset correlation values which in turn leads to a better protection of the financial system.  

\subsection{Financial Inclusion}
The results may also have a societal impact that affect borrowers in developed and developing countries in different ways.
In the former case, people who are joining the financial market for the first time, such as young people and immigrants, face troubles when applying for loans because they do not have a credit history.
Instead, they need to spend time and effort to build their credit history before financial institutions can assess whether they are creditworthy.

In developing countries where historical financial data is often nonexistent, the impact is even greater.
As reported by the World Bank, over two billion adults worldwide do not have a basic account which makes up more than $20\%$ of the adult population in some countries \citep{worldBank}.
The benefits of behavior-based microfinance in these countries are evident, as having access to small credits has a social impact on communities, helping to fight poverty and enhancing economic development \citep{copestake2007mainstreaming}.
In contrast to the lack of banking history, the high(er) availability of call data
in these countries provides an alternative for credit scoring, hereby facilitating credit access to a wider segment of the population.
According to the results, features extracted from these untraditional data sources are good predictors of credit behavior (e.g., models $B$, $G$ and $H$).
In addition, the numerous smartphone applications that are already being deployed in some developing countries are a prime example of the success of these methods. 
They offer immediate small loans, that are repaid within a short period of between three weeks and six months and have lower interest rates, ranging between $6\%$ and $12\%$ as opposed to the $25\%$ interest rate in traditional microlending \citep{dwoskin2015lending}.

\subsection{Privacy and Ethical Concerns}
The results of this study are furthermore affected by privacy regulations because the implementation of some of the models depends on different parties sharing the data.
Since there are no worldwide applicable standards for data-sharing of that kind, we illustrate how this might occur by studying the reality of the US and the EU in what follows.

In the US, there is no single federal law regulating data transfer between affiliates. The transfer of financial information between a bank and a telco is protected under the Gramm-Leach-Bliley Act \citep{Cuaresma2002}. This legislation allows transfer of personally identifiable information originating from a financial service provider to a third party if the parties design a contract that disallows disclosure and use of information outside the project. In general, such a contractual framework should satisfy most other pieces of regulation that might indirectly apply to the sharing of data in the other direction (i.e., from the telco to the bank or credit bureau).

In the European Union, in contrast, there is a strong body of legislation regulating data sharing. Given that CDR data are a form of communication, and the objective of the model is to process it along with banking data in an automated way, two pieces of legislation apply: Regulation 2016/679 \citep{EU_GDPR} regarding the protection of privacy for natural persons, best known as the General Data Protection Regulation (GDPR), and the ``ePrivacy Regulation'' \citep{EU_ePrivacy} regarding the processing of personal data in the electronics communications sector.

The ePrivacy Regulation deals with if, and how, communications data at a disaggregated level can be used. Article 30 in particular mandates that a service such as a financial score, which is not only for billing or providing the mobile service, is a ``value added service'', and thus requires explicit authorization from the user. This authorization might be given in the contract, for example, or ex post to the signing of the contract via electronic authorization. In case none of these provisions can be set in place, then the sharing of CDR data cannot occur unless the data is anonymized. 

The key challenge is how to make the data available to the other party, so defaulters can be correctly identified. Fortunately, there are methods that can provide privacy-preserving data linkage \citep{Clifton2004} that can be followed in order to join the data securely without compromising the individual on either side of the sharing process. Methods such as Privacy Preserving Probabilistic Record Linkage \citep[P3RL][]{Schmidlin2015}, that are in use in the medical sciences, allow secure data-sharing between partners. The secure transfer of data is also very simple to satisfy, following proper encryption and secure access protocols. The GDPR has additional provisions on data storage, forcing companies to store data only for the time necessary to provide the service, so the party receiving the linked data only for the purposes of model development must ensure proper disposal of the data after the model development. Finally, note that the model itself is considered aggregated data. Article 22 of the GDPR allows the safe use of statistical models when these are required to establish a contract with the counterparty (the financial institution), which is the case when a loan is granted.

There is as well an ethical concern in using data that depends on the social network of the borrowers to restrict funding to them. This is of course not a practice that should be recommended from the results of this model, as it would constitute unfair discrimination. However, when borrowers do not have any past behavior information that allows institutions to make a decision, or they have not accumulated enough additional information to profile them correctly, then CDR information can clearly contribute to increase financial inclusion. Thus, we propose that the use of this data be done in strict positive terms. This can be easily done when constructing a credit score: it is common practice to discretize continuous variables and give a score based on the Weight of Evidence for each of the segments \citep{scheule2016credit}. An ethical use of this information would simply assign the neutral score to those segment which would unfairly punish the borrowers, leaving the positive segments that would provide easier access to funds.

\section{Conclusion}\label{sec:conclusion}
This study presents the statistical and economic advantages of exploiting Big Data and social network analytics for credit scoring applications.
We use phone call logs are used to build call networks and social network analytics applied to enhance the performance of models that predict creditworthiness of credit cards applicants.
We do this from both a statistical and profit perspective and demonstrate how incorporating telco data has the potential of increasing the Value of credit scoring models. 
Furthermore, we identify which features are most important for this predictive task, both in terms of statistical performance and profit.
According to the results, models that are built with features that represent calling behavior perform best, both when performance is measured in AUC and profit.
We also show that these features dominate other features in terms of importance.
This is an interesting result because it means that how people use their phones can be used as the sole data source when deciding whether they should be given a loan or not.
Thus we propose that the data should be used in strict positive terms, to facilitate financial inclusion for people that lack enough information for correct profiling.

The main limitation of our this is the data itself.
The scorecards that were built are for the applications of credit cards, and it is unclear how the results would generalize for other types of credits such as microloans or mortgages.
In industry, numerous applications for granting microloans via smartphones by analyzing user's behavior exist.
According to various reports, behavioral features are important in these applications as well, but that is difficult to verify without published scientific results.
Similar data could be obtained from peer-to-peer lending platforms, or through agreements between telcos and banks/credit bureaus, where there is access to both default status of users as well as behavioral features.
Behavioral data similar to the mobile phone data shown in this work could also be gathered from social media platforms such as Twitter.
The data in this study originates from a single country where a telco and a bank have a special agreement to share the data.
Therefore, an analysis of similar data from other countries or data for other types of credits would strengthen the external validity of the presented results.
In practice, lenders use credit bureau variables, such as FICO scores, when assessing creditworthiness, and unfortunately they were not available for these analyses, but would be an interesting extension of our work. 

It is already clear that the mobile phone data used in this study is big in the sense of `Volume', `Velocity', `Veracity' and `Variety'.
Our analysis of the data and the resulting well-performing models show that it also has a positive effect for financial inclusion and on model profit, and as such is also important for `Value': the fifth V of Big Data!

\section*{Acknowledgments}
The authors would like to thank Ariel Berenstein for his contribution to this research.

The authors acknowledge the support of a large Belgian bank that wishes to remain anonymous.

\footnotesize{
\bibliography{credit_bib}

\begin{thebibliography}{52}
\providecommand{\natexlab}[1]{#1}
\providecommand{\url}[1]{\texttt{#1}}
\providecommand{\href}[2]{#2}
\providecommand{\path}[1]{#1}
\providecommand{\eprint}[1]{\href{http://arxiv.org/abs/#1}{\path{#1}}}
\providecommand{\DOIprefix}{doi:}
\providecommand{\ArXivprefix}{arXiv:}
\providecommand{\URLprefix}{URL: }
\providecommand{\Pubmedprefix}{pmid:}
\providecommand{\doi}[1]{\href{http://dx.doi.org/#1}{\path{#1}}}
\providecommand{\Pubmed}[1]{\href{pmid:#1}{\path{#1}}}
\providecommand{\BIBand}{and}
\providecommand{\bibinfo}[2]{#2}
\ifx\xfnm\undefined \def\xfnm[#1]{\unskip,\space#1}\fi
\bibitem[{Thomas(2000)}]{thomas2000survey}
\bibinfo{author}{Thomas\xfnm[ L.C.]}.
\newblock \bibinfo{title}{A survey of credit and behavioural scoring:
  forecasting financial risk of lending to consumers}.
\newblock \bibinfo{journal}{International journal of forecasting}
  \bibinfo{year}{2000};\bibinfo{volume}{16}(\bibinfo{number}{2}):\bibinfo{pages}{149--172}.
\bibitem[{Baesens et~al.(2016)Baesens, Scheule and
  R{\"o}sch}]{scheule2016credit}
\bibinfo{author}{Baesens\xfnm[ B.]}, \bibinfo{author}{Scheule\xfnm[ H.]},
  \bibinfo{author}{R{\"o}sch\xfnm[ D.]}.
\newblock \bibinfo{title}{Credit Risk Analytics: Measurement Techniques,
  Applications, and Examples in SAS}.
\newblock \bibinfo{publisher}{John Wiley \& Sons}; \bibinfo{year}{2016}.
\bibitem[{Altman(1968)}]{altman1968financial}
\bibinfo{author}{Altman\xfnm[ E.I.]}.
\newblock \bibinfo{title}{Financial ratios, discriminant analysis and the
  prediction of corporate bankruptcy}.
\newblock \bibinfo{journal}{The journal of finance}
  \bibinfo{year}{1968};\bibinfo{volume}{23}(\bibinfo{number}{4}):\bibinfo{pages}{589--609}.
\bibitem[{Baesens et~al.(2003)Baesens, Van~Gestel, Viaene, Stepanova, Suykens
  and Vanthienen}]{baesens2003benchmarking}
\bibinfo{author}{Baesens\xfnm[ B.]}, \bibinfo{author}{Van~Gestel\xfnm[ T.]},
  \bibinfo{author}{Viaene\xfnm[ S.]}, \bibinfo{author}{Stepanova\xfnm[ M.]},
  \bibinfo{author}{Suykens\xfnm[ J.]}, \bibinfo{author}{Vanthienen\xfnm[ J.]}.
\newblock \bibinfo{title}{Benchmarking state-of-the-art classification
  algorithms for credit scoring}.
\newblock \bibinfo{journal}{Journal of the operational research society}
  \bibinfo{year}{2003};\bibinfo{volume}{54}(\bibinfo{number}{6}):\bibinfo{pages}{627--635}.
\bibitem[{Lessmann et~al.(2015)Lessmann, Baesens, Seow and
  Thomas}]{lessmann2015benchmarking}
\bibinfo{author}{Lessmann\xfnm[ S.]}, \bibinfo{author}{Baesens\xfnm[ B.]},
  \bibinfo{author}{Seow\xfnm[ H.V.]}, \bibinfo{author}{Thomas\xfnm[ L.C.]}.
\newblock \bibinfo{title}{Benchmarking state-of-the-art classification
  algorithms for credit scoring: An update of research}.
\newblock \bibinfo{journal}{Eur J Oper Res}
  \bibinfo{year}{2015};\bibinfo{volume}{247}(\bibinfo{number}{1}):\bibinfo{pages}{124--136}.
\bibitem[{{World Bank}(2011)}]{worldbank2011}
\bibinfo{author}{{World Bank}\xfnm[]}.
\newblock \bibinfo{title}{General principles for credit reporting}.
\newblock \bibinfo{year}{2011}.
\newblock \URLprefix
  \url{http://documents.worldbank.org/curated/en/662161468147557554/pdf/70193-2014-CR-General-Principles-Web-Ready.pdf}.
\bibitem[{Barron and Staten(2003)}]{barron2003value}
\bibinfo{author}{Barron\xfnm[ J.M.]}, \bibinfo{author}{Staten\xfnm[ M.]}.
\newblock \bibinfo{title}{The value of comprehensive credit reports: Lessons
  from the us experience}.
\newblock \bibinfo{journal}{Credit reporting systems and the international
  economy} \bibinfo{year}{2003};\bibinfo{volume}{8}:\bibinfo{pages}{273--310}.
\bibitem[{Agarwal and Dhar(2014)}]{agarwal2014editorial}
\bibinfo{author}{Agarwal\xfnm[ R.]}, \bibinfo{author}{Dhar\xfnm[ V.]}.
\newblock \bibinfo{title}{Editorial—-{Big} data, data science, and analytics:
  The opportunity and challenge for {IS} research}.
\newblock \bibinfo{year}{2014}.
\bibitem[{Baesens et~al.(2014)Baesens, Bapna, Marsden, Vanthienen and
  Zhao}]{baesens2014transformational}
\bibinfo{author}{Baesens\xfnm[ B.]}, \bibinfo{author}{Bapna\xfnm[ R.]},
  \bibinfo{author}{Marsden\xfnm[ J.R.]}, \bibinfo{author}{Vanthienen\xfnm[
  J.]}, \bibinfo{author}{Zhao\xfnm[ J.L.]}.
\newblock \bibinfo{title}{Transformational issues of big data and analytics in
  networked business}.
\newblock \bibinfo{journal}{MIS quarterly}
  \bibinfo{year}{2014};\bibinfo{volume}{38}(\bibinfo{number}{2}):\bibinfo{pages}{629--631}.
\bibitem[{Sundararajan et~al.(2013)Sundararajan, Provost, Oestreicher-Singer
  and Aral}]{sundararajan2013research}
\bibinfo{author}{Sundararajan\xfnm[ A.]}, \bibinfo{author}{Provost\xfnm[ F.]},
  \bibinfo{author}{Oestreicher-Singer\xfnm[ G.]}, \bibinfo{author}{Aral\xfnm[
  S.]}.
\newblock \bibinfo{title}{Research commentary —- information in digital,
  economic, and social networks}.
\newblock \bibinfo{journal}{Inform Systems Res}
  \bibinfo{year}{2013};\bibinfo{volume}{24}(\bibinfo{number}{4}):\bibinfo{pages}{883--905}.
\bibitem[{Newman(2010)}]{newman2010networks}
\bibinfo{author}{Newman\xfnm[ M.]}.
\newblock \bibinfo{title}{Networks: an introduction}.
\newblock \bibinfo{publisher}{OUP Oxford}; \bibinfo{year}{2010}.
\bibitem[{Lee et~al.(2016)Lee, Qiu and Whinston}]{lee2016friend}
\bibinfo{author}{Lee\xfnm[ G.M.]}, \bibinfo{author}{Qiu\xfnm[ L.]},
  \bibinfo{author}{Whinston\xfnm[ A.B.]}.
\newblock \bibinfo{title}{A friend like me: modeling network formation in a
  location-based social network}.
\newblock \bibinfo{journal}{Journal of Management Information Systems}
  \bibinfo{year}{2016};\bibinfo{volume}{33}(\bibinfo{number}{4}):\bibinfo{pages}{1008--1033}.
\bibitem[{Aral and Walker(2014)}]{aral2014tie}
\bibinfo{author}{Aral\xfnm[ S.]}, \bibinfo{author}{Walker\xfnm[ D.]}.
\newblock \bibinfo{title}{Tie strength, embeddedness, and social influence: A
  large-scale networked experiment}.
\newblock \bibinfo{journal}{Management Sci}
  \bibinfo{year}{2014};\bibinfo{volume}{60}(\bibinfo{number}{6}):\bibinfo{pages}{1352--1370}.
\bibitem[{Macskassy and Provost(2007)}]{macskassy2007classification}
\bibinfo{author}{Macskassy\xfnm[ S.A.]}, \bibinfo{author}{Provost\xfnm[ F.]}.
\newblock \bibinfo{title}{Classification in networked data: A toolkit and a
  univariate case study}.
\newblock \bibinfo{journal}{Journal of Machine Learning Research}
  \bibinfo{year}{2007};\bibinfo{volume}{8}(\bibinfo{number}{May}):\bibinfo{pages}{935--983}.
\bibitem[{Verbeke et~al.(2014)Verbeke, Martens and Baesens}]{verbeke2014social}
\bibinfo{author}{Verbeke\xfnm[ W.]}, \bibinfo{author}{Martens\xfnm[ D.]},
  \bibinfo{author}{Baesens\xfnm[ B.]}.
\newblock \bibinfo{title}{Social network analysis for customer churn
  prediction}.
\newblock \bibinfo{journal}{Applied Soft Computing}
  \bibinfo{year}{2014};\bibinfo{volume}{14}:\bibinfo{pages}{431--446}.
\bibitem[{Van~Vlasselaer et~al.(2015)Van~Vlasselaer, Bravo, Caelen,
  Eliassi-Rad, Akoglu, Snoeck et~al.}]{van2015apate}
\bibinfo{author}{Van~Vlasselaer\xfnm[ V.]}, \bibinfo{author}{Bravo\xfnm[ C.]},
  \bibinfo{author}{Caelen\xfnm[ O.]}, \bibinfo{author}{Eliassi-Rad\xfnm[ T.]},
  \bibinfo{author}{Akoglu\xfnm[ L.]}, \bibinfo{author}{Snoeck\xfnm[ M.]},
  et~al.
\newblock \bibinfo{title}{Apate: A novel approach for automated credit card
  transaction fraud detection using network-based extensions}.
\newblock \bibinfo{journal}{Decision Support Systems}
  \bibinfo{year}{2015};\bibinfo{volume}{75}:\bibinfo{pages}{38--48}.
\bibitem[{Gordy(2003)}]{gordy2003risk}
\bibinfo{author}{Gordy\xfnm[ M.B.]}.
\newblock \bibinfo{title}{A risk-factor model foundation for ratings-based bank
  capital rules}.
\newblock \bibinfo{journal}{Journal of financial intermediation}
  \bibinfo{year}{2003};\bibinfo{volume}{12}(\bibinfo{number}{3}):\bibinfo{pages}{199--232}.
\bibitem[{Calabrese et~al.(2017)Calabrese, Andreeva and
  Ansell}]{calabrese2017birds}
\bibinfo{author}{Calabrese\xfnm[ R.]}, \bibinfo{author}{Andreeva\xfnm[ G.]},
  \bibinfo{author}{Ansell\xfnm[ J.]}.
\newblock \bibinfo{title}{“{B}irds of a feather” fail together: Exploring
  the nature of dependency in {SME} defaults}.
\newblock \bibinfo{journal}{Risk Analysis} \bibinfo{year}{2017};.
\bibitem[{Lin et~al.(2013)Lin, Prabhala and Viswanathan}]{lin2013judging}
\bibinfo{author}{Lin\xfnm[ M.]}, \bibinfo{author}{Prabhala\xfnm[ N.R.]},
  \bibinfo{author}{Viswanathan\xfnm[ S.]}.
\newblock \bibinfo{title}{Judging borrowers by the company they keep:
  friendship networks and information asymmetry in online peer-to-peer
  lending}.
\newblock \bibinfo{journal}{Management Sci}
  \bibinfo{year}{2013};\bibinfo{volume}{59}(\bibinfo{number}{1}):\bibinfo{pages}{17--35}.
\bibitem[{Freedman and Jin(2014)}]{freedman2014information}
\bibinfo{author}{Freedman\xfnm[ S.]}, \bibinfo{author}{Jin\xfnm[ G.Z.]}.
\newblock \bibinfo{title}{The information value of online social networks:
  lessons from peer-to-peer lending}.
\newblock \bibinfo{type}{Tech. Rep.}; National Bureau of Economic Research;
  \bibinfo{year}{2014}.
\bibitem[{Wei et~al.(2015)Wei, Yildirim, Van~den Bulte and
  Dellarocas}]{wei2015credit}
\bibinfo{author}{Wei\xfnm[ Y.]}, \bibinfo{author}{Yildirim\xfnm[ P.]},
  \bibinfo{author}{Van~den Bulte\xfnm[ C.]}, \bibinfo{author}{Dellarocas\xfnm[
  C.]}.
\newblock \bibinfo{title}{Credit scoring with social network data}.
\newblock \bibinfo{journal}{Marketing Sci}
  \bibinfo{year}{2015};\bibinfo{volume}{35}(\bibinfo{number}{2}):\bibinfo{pages}{234--258}.
\bibitem[{De~Cnudde et~al.(2018)De~Cnudde, Moeyersoms, Stankova, Tobback,
  Javaly and Martens}]{de2018does}
\bibinfo{author}{De~Cnudde\xfnm[ S.]}, \bibinfo{author}{Moeyersoms\xfnm[ J.]},
  \bibinfo{author}{Stankova\xfnm[ M.]}, \bibinfo{author}{Tobback\xfnm[ E.]},
  \bibinfo{author}{Javaly\xfnm[ V.]}, \bibinfo{author}{Martens\xfnm[ D.]}.
\newblock \bibinfo{title}{What does your facebook profile reveal about your
  creditworthiness? using alternative data for microfinance}.
\newblock \bibinfo{journal}{Journal of the Operational Research Society}
  \bibinfo{year}{2018};:\bibinfo{pages}{1--11}.
\bibitem[{Kharif(2016)}]{kharif2016no}
\bibinfo{author}{Kharif\xfnm[ O.]}.
\newblock \bibinfo{title}{No credit history? no problem. lenders are looking at
  your phone data}.
\newblock \bibinfo{year}{2016}.
\newblock \URLprefix
  \url{https://www.bloomberg.com/news/articles/2016-11-25/no-credit-history-no-problem-lenders-now-peering-at-phone-data}.
\bibitem[{Chin and Wong(2016)}]{chin2016china}
\bibinfo{author}{Chin\xfnm[ J.]}, \bibinfo{author}{Wong\xfnm[ G.]}.
\newblock \bibinfo{title}{China's new tool for social control: A credit rating
  for everything}.
\newblock \bibinfo{journal}{The Wall Street Journal} \bibinfo{year}{2016};.
\bibitem[{Dwoskin(2015)}]{dwoskin2015lending}
\bibinfo{author}{Dwoskin\xfnm[ E.]}.
\newblock \bibinfo{title}{Lending startups look at borrowers’ phone usage to
  assess creditworthiness}.
\newblock \bibinfo{journal}{The Wall Street Journal} \bibinfo{year}{2015};.
\bibitem[{Naboulsi et~al.(2016)Naboulsi, Fiore, Ribot and
  Stanica}]{naboulsi2016large}
\bibinfo{author}{Naboulsi\xfnm[ D.]}, \bibinfo{author}{Fiore\xfnm[ M.]},
  \bibinfo{author}{Ribot\xfnm[ S.]}, \bibinfo{author}{Stanica\xfnm[ R.]}.
\newblock \bibinfo{title}{Large-scale mobile traffic analysis: a survey}.
\newblock \bibinfo{journal}{IEEE Communications Surveys \& Tutorials}
  \bibinfo{year}{2016};\bibinfo{volume}{18}(\bibinfo{number}{1}):\bibinfo{pages}{124--161}.
\bibitem[{Blondel et~al.(2008)Blondel, Guillaume, Lambiotte and
  Lefebvre}]{blondel2008fast}
\bibinfo{author}{Blondel\xfnm[ V.D.]}, \bibinfo{author}{Guillaume\xfnm[ J.L.]},
  \bibinfo{author}{Lambiotte\xfnm[ R.]}, \bibinfo{author}{Lefebvre\xfnm[ E.]}.
\newblock \bibinfo{title}{Fast unfolding of communities in large networks}.
\newblock \bibinfo{journal}{Journal of statistical mechanics: theory and
  experiment}
  \bibinfo{year}{2008};\bibinfo{volume}{2008}(\bibinfo{number}{10}):\bibinfo{pages}{P10008}.
\bibitem[{Onnela et~al.(2011)Onnela, Arbesman, Gonz{\'a}lez, Barab{\'a}si and
  Christakis}]{onnela2011geographic}
\bibinfo{author}{Onnela\xfnm[ J.P.]}, \bibinfo{author}{Arbesman\xfnm[ S.]},
  \bibinfo{author}{Gonz{\'a}lez\xfnm[ M.C.]},
  \bibinfo{author}{Barab{\'a}si\xfnm[ A.L.]}, \bibinfo{author}{Christakis\xfnm[
  N.A.]}.
\newblock \bibinfo{title}{Geographic constraints on social network groups}.
\newblock \bibinfo{journal}{PLoS one}
  \bibinfo{year}{2011};\bibinfo{volume}{6}(\bibinfo{number}{4}):\bibinfo{pages}{e16939}.
\bibitem[{Wesolowski et~al.(2012)Wesolowski, Eagle, Tatem, Smith, Noor, Snow
  et~al.}]{wesolowski2012quantifying}
\bibinfo{author}{Wesolowski\xfnm[ A.]}, \bibinfo{author}{Eagle\xfnm[ N.]},
  \bibinfo{author}{Tatem\xfnm[ A.J.]}, \bibinfo{author}{Smith\xfnm[ D.L.]},
  \bibinfo{author}{Noor\xfnm[ A.M.]}, \bibinfo{author}{Snow\xfnm[ R.W.]},
  et~al.
\newblock \bibinfo{title}{Quantifying the impact of human mobility on malaria}.
\newblock \bibinfo{journal}{Science}
  \bibinfo{year}{2012};\bibinfo{volume}{338}(\bibinfo{number}{6104}):\bibinfo{pages}{267--270}.
\bibitem[{Sarraute et~al.(2014)Sarraute, Blanc and Burroni}]{sarraute2014study}
\bibinfo{author}{Sarraute\xfnm[ C.]}, \bibinfo{author}{Blanc\xfnm[ P.]},
  \bibinfo{author}{Burroni\xfnm[ J.]}.
\newblock \bibinfo{title}{A study of age and gender seen through mobile phone
  usage patterns in mexico}.
\newblock In: \bibinfo{booktitle}{Advances in Social Networks Analysis and
  Mining (ASONAM), 2014 IEEE/ACM International Conference on}.
  \bibinfo{organization}{IEEE}; \bibinfo{year}{2014}, p.
  \bibinfo{pages}{836--843}.
\bibitem[{Leo et~al.(2016)Leo, Fleury, Alvarez-Hamelin, Sarraute and
  Karsai}]{leo2016socioeconomic}
\bibinfo{author}{Leo\xfnm[ Y.]}, \bibinfo{author}{Fleury\xfnm[ E.]},
  \bibinfo{author}{Alvarez-Hamelin\xfnm[ J.I.]},
  \bibinfo{author}{Sarraute\xfnm[ C.]}, \bibinfo{author}{Karsai\xfnm[ M.]}.
\newblock \bibinfo{title}{Socioeconomic correlations and stratification in
  social-communication networks}.
\newblock \bibinfo{journal}{Journal of The Royal Society Interface}
  \bibinfo{year}{2016};\bibinfo{volume}{13}(\bibinfo{number}{125}):\bibinfo{pages}{20160598}.
\bibitem[{Haenlein(2011)}]{haenlein2011social}
\bibinfo{author}{Haenlein\xfnm[ M.]}.
\newblock \bibinfo{title}{A social network analysis of customer-level revenue
  distribution}.
\newblock \bibinfo{journal}{Marketing letters}
  \bibinfo{year}{2011};\bibinfo{volume}{22}(\bibinfo{number}{1}):\bibinfo{pages}{15--29}.
\bibitem[{Van~Vlasselaer et~al.(2016)Van~Vlasselaer, Eliassi-Rad, Akoglu,
  Snoeck and Baesens}]{van2016gotcha}
\bibinfo{author}{Van~Vlasselaer\xfnm[ V.]}, \bibinfo{author}{Eliassi-Rad\xfnm[
  T.]}, \bibinfo{author}{Akoglu\xfnm[ L.]}, \bibinfo{author}{Snoeck\xfnm[ M.]},
  \bibinfo{author}{Baesens\xfnm[ B.]}.
\newblock \bibinfo{title}{Gotcha! network-based fraud detection for social
  security fraud}.
\newblock \bibinfo{journal}{Management Science}
  \bibinfo{year}{2016};\bibinfo{volume}{63}(\bibinfo{number}{9}):\bibinfo{pages}{3090--3110}.
\bibitem[{Lu and Getoor(2003)}]{lu2003link}
\bibinfo{author}{Lu\xfnm[ Q.]}, \bibinfo{author}{Getoor\xfnm[ L.]}.
\newblock \bibinfo{title}{Link-based classification}.
\newblock In: \bibinfo{booktitle}{ICML}; vol.~\bibinfo{volume}{3}.
  \bibinfo{year}{2003}, p. \bibinfo{pages}{496--503}.
\bibitem[{{\'O}skarsd{\'o}ttir et~al.(2017){\'O}skarsd{\'o}ttir, Bravo,
  Verbeke, Sarraute, Baesens and Vanthienen}]{oskarsdottir2017social}
\bibinfo{author}{{\'O}skarsd{\'o}ttir\xfnm[ M.]}, \bibinfo{author}{Bravo\xfnm[
  C.]}, \bibinfo{author}{Verbeke\xfnm[ W.]}, \bibinfo{author}{Sarraute\xfnm[
  C.]}, \bibinfo{author}{Baesens\xfnm[ B.]}, \bibinfo{author}{Vanthienen\xfnm[
  J.]}.
\newblock \bibinfo{title}{Social network analytics for churn prediction in
  telco: Model building, evaluation and network architecture}.
\newblock \bibinfo{journal}{Expert Systems with Applications}
  \bibinfo{year}{2017};\bibinfo{volume}{85}:\bibinfo{pages}{204--220}.
\bibitem[{Page et~al.(1999)Page, Brin, Motwani and Winograd}]{page1999pagerank}
\bibinfo{author}{Page\xfnm[ L.]}, \bibinfo{author}{Brin\xfnm[ S.]},
  \bibinfo{author}{Motwani\xfnm[ R.]}, \bibinfo{author}{Winograd\xfnm[ T.]}.
\newblock \bibinfo{title}{The pagerank citation ranking: Bringing order to the
  web.}
\newblock \bibinfo{type}{Tech. Rep.}; Stanford InfoLab; \bibinfo{year}{1999}.
\bibitem[{Dasgupta et~al.(2008)Dasgupta, Singh, Viswanathan, Chakraborty,
  Mukherjea, Nanavati et~al.}]{dasgupta2008social}
\bibinfo{author}{Dasgupta\xfnm[ K.]}, \bibinfo{author}{Singh\xfnm[ R.]},
  \bibinfo{author}{Viswanathan\xfnm[ B.]}, \bibinfo{author}{Chakraborty\xfnm[
  D.]}, \bibinfo{author}{Mukherjea\xfnm[ S.]}, \bibinfo{author}{Nanavati\xfnm[
  A.A.]}, et~al.
\newblock \bibinfo{title}{Social ties and their relevance to churn in mobile
  telecom networks}.
\newblock In: \bibinfo{booktitle}{Proceedings of the 11th international
  conference on Extending database technology: Advances in database
  technology}. \bibinfo{organization}{ACM}; \bibinfo{year}{2008}, p.
  \bibinfo{pages}{668--677}.
\bibitem[{Backiel et~al.(2015)Backiel, Verbinnen, Baesens and
  Claeskens}]{backiel2015combining}
\bibinfo{author}{Backiel\xfnm[ A.]}, \bibinfo{author}{Verbinnen\xfnm[ Y.]},
  \bibinfo{author}{Baesens\xfnm[ B.]}, \bibinfo{author}{Claeskens\xfnm[ G.]}.
\newblock \bibinfo{title}{Combining local and social network classifiers to
  improve churn prediction}.
\newblock In: \bibinfo{booktitle}{Proceedings of the 2015 IEEE/ACM
  International Conference on Advances in Social Networks Analysis and Mining
  2015}. \bibinfo{organization}{ACM}; \bibinfo{year}{2015}, p.
  \bibinfo{pages}{651--658}.
\bibitem[{Verbraken et~al.(2014)Verbraken, Bravo, Weber and
  Baesens}]{verbraken2014development}
\bibinfo{author}{Verbraken\xfnm[ T.]}, \bibinfo{author}{Bravo\xfnm[ C.]},
  \bibinfo{author}{Weber\xfnm[ R.]}, \bibinfo{author}{Baesens\xfnm[ B.]}.
\newblock \bibinfo{title}{Development and application of consumer credit
  scoring models using profit-based classification measures}.
\newblock \bibinfo{journal}{Eur J Oper Res}
  \bibinfo{year}{2014};\bibinfo{volume}{238}(\bibinfo{number}{2}):\bibinfo{pages}{505--513}.
\bibitem[{Verbraken et~al.(2013)Verbraken, Verbeke and
  Baesens}]{verbraken2013novel}
\bibinfo{author}{Verbraken\xfnm[ T.]}, \bibinfo{author}{Verbeke\xfnm[ W.]},
  \bibinfo{author}{Baesens\xfnm[ B.]}.
\newblock \bibinfo{title}{A novel profit maximizing metric for measuring
  classification performance of customer churn prediction models}.
\newblock \bibinfo{journal}{IEEE transactions on knowledge and data
  engineering}
  \bibinfo{year}{2013};\bibinfo{volume}{25}(\bibinfo{number}{5}):\bibinfo{pages}{961--973}.
\bibitem[{Bravo et~al.(2013)Bravo, Maldonado and Weber}]{bravo2013granting}
\bibinfo{author}{Bravo\xfnm[ C.]}, \bibinfo{author}{Maldonado\xfnm[ S.]},
  \bibinfo{author}{Weber\xfnm[ R.]}.
\newblock \bibinfo{title}{Granting and managing loans for micro-entrepreneurs:
  New developments and practical experiences}.
\newblock \bibinfo{journal}{European Journal of Operational Research}
  \bibinfo{year}{2013};\bibinfo{volume}{227}(\bibinfo{number}{2}):\bibinfo{pages}{358--366}.
\bibitem[{Singh et~al.(2015)Singh, Bozkaya and Pentland}]{singh2015money}
\bibinfo{author}{Singh\xfnm[ V.K.]}, \bibinfo{author}{Bozkaya\xfnm[ B.]},
  \bibinfo{author}{Pentland\xfnm[ A.]}.
\newblock \bibinfo{title}{Money walks: implicit mobility behavior and financial
  well-being}.
\newblock \bibinfo{journal}{PloS one}
  \bibinfo{year}{2015};\bibinfo{volume}{10}(\bibinfo{number}{8}):\bibinfo{pages}{e0136628}.
\bibitem[{Baesens(2014)}]{baesens2014analytics}
\bibinfo{author}{Baesens\xfnm[ B.]}.
\newblock \bibinfo{title}{Analytics in a big data world: The essential guide to
  data science and its applications}.
\newblock \bibinfo{publisher}{John Wiley \& Sons}; \bibinfo{year}{2014}.
\bibitem[{Baesens et~al.(2015)Baesens, Van~Vlasselaer and
  Verbeke}]{baesens2015fraud}
\bibinfo{author}{Baesens\xfnm[ B.]}, \bibinfo{author}{Van~Vlasselaer\xfnm[
  V.]}, \bibinfo{author}{Verbeke\xfnm[ W.]}.
\newblock \bibinfo{title}{Fraud analytics using descriptive, predictive, and
  social network techniques: a guide to data science for fraud detection}.
\newblock \bibinfo{publisher}{John Wiley \& Sons}; \bibinfo{year}{2015}.
\bibitem[{DeLong et~al.(1988)DeLong, DeLong and
  Clarke-Pearson}]{delong1988comparing}
\bibinfo{author}{DeLong\xfnm[ E.R.]}, \bibinfo{author}{DeLong\xfnm[ D.M.]},
  \bibinfo{author}{Clarke-Pearson\xfnm[ D.L.]}.
\newblock \bibinfo{title}{Comparing the areas under two or more correlated
  receiver operating characteristic curves: a nonparametric approach}.
\newblock \bibinfo{journal}{Biometrics}
  \bibinfo{year}{1988};:\bibinfo{pages}{837--845}.
\bibitem[{{World Bank}(2017)}]{worldBank}
\bibinfo{author}{{World Bank}\xfnm[]}.
\newblock \bibinfo{title}{Financial inclusion}.
\newblock \bibinfo{year}{2017}.
\newblock \URLprefix
  \url{http://www.worldbank.org/en/topic/financialinclusion/overview#1}.
\bibitem[{Copestake(2007)}]{copestake2007mainstreaming}
\bibinfo{author}{Copestake\xfnm[ J.]}.
\newblock \bibinfo{title}{Mainstreaming microfinance: Social performance
  management or mission drift?}
\newblock \bibinfo{journal}{World Development}
  \bibinfo{year}{2007};\bibinfo{volume}{35}(\bibinfo{number}{10}):\bibinfo{pages}{1721--1738}.
\bibitem[{Cuaresma(2002)}]{Cuaresma2002}
\bibinfo{author}{Cuaresma\xfnm[ J.C.]}.
\newblock \bibinfo{title}{The {G}ramm-{L}each-{B}liley {A}ct}.
\newblock \bibinfo{journal}{Berkeley Technology Law Journal}
  \bibinfo{year}{2002};\bibinfo{volume}{17}(\bibinfo{number}{1}):\bibinfo{pages}{497--517}.
\bibitem[{{European Union}(2016)}]{EU_GDPR}
\bibinfo{author}{{European Union}\xfnm[]}.
\newblock \bibinfo{title}{Protection of natural persons with regard to the
  processing of personal data and on the free movement of such data, and
  repealing {D}irective 95/46/{EC} ({General Data Protection Regulation})}.
\newblock \bibinfo{howpublished}{Legislation}; \bibinfo{year}{2016}.
\bibitem[{{European Union}(2002)}]{EU_ePrivacy}
\bibinfo{author}{{European Union}\xfnm[]}.
\newblock \bibinfo{title}{Directive 2002/58/ec concerning the processing of
  personal data and the protection of privacy in the electronic communications
  sector}.
\newblock \bibinfo{howpublished}{Legislation}; \bibinfo{year}{2002}.
\bibitem[{Clifton et~al.(2004)Clifton, Kantarcio\v{g}lu, Doan, Schadow, Vaidya,
  Elmagarmid et~al.}]{Clifton2004}
\bibinfo{author}{Clifton\xfnm[ C.]}, \bibinfo{author}{Kantarcio\v{g}lu\xfnm[
  M.]}, \bibinfo{author}{Doan\xfnm[ A.]}, \bibinfo{author}{Schadow\xfnm[ G.]},
  \bibinfo{author}{Vaidya\xfnm[ J.]}, \bibinfo{author}{Elmagarmid\xfnm[ A.]},
  et~al.
\newblock \bibinfo{title}{Privacy-preserving data integration and sharing}.
\newblock In: \bibinfo{booktitle}{Proceedings of the 9th ACM SIGMOD workshop on
  Research issues in data mining and knowledge discovery}.
  \bibinfo{organization}{ACM}; \bibinfo{year}{2004}, p.
  \bibinfo{pages}{19--26}.
\bibitem[{Schmidlin et~al.(2015)Schmidlin, Clough-Gorr and
  Spoerri}]{Schmidlin2015}
\bibinfo{author}{Schmidlin\xfnm[ K.]}, \bibinfo{author}{Clough-Gorr\xfnm[
  K.M.]}, \bibinfo{author}{Spoerri\xfnm[ A.]}.
\newblock \bibinfo{title}{Privacy preserving probabilistic record linkage
  (p3rl): a novel method for linking existing health-related data and
  maintaining participant confidentiality}.
\newblock \bibinfo{journal}{BMC medical research methodology}
  \bibinfo{year}{2015};\bibinfo{volume}{15}(\bibinfo{number}{1}):\bibinfo{pages}{46}.

\end{thebibliography}
}
\end{document}